\documentclass[12pt]{iopart}
\usepackage{graphicx}
\usepackage{iopams}  
\begin{document}

\title[Systematics at the micro and nano length scale]{Modeling the systematic behavior at the micro and nano length scale}

\author{Danilo Quagliotti}

\address{Technical University of Denmark, Department of Mechanical Engineering, Produktionstorvet 427A, Kgs. Lyngby, Denmark}
\ead{danqua@mek.dtu.dk}
\vspace{10pt}
\begin{indented}
\item[]July 2021
\end{indented}

\begin{abstract}
The brisk progression of the industrial \textit{digital} innovation, leading to high degree of automation and \textit{big data} transfer in manufacturing technologies, demands continuous development of appropriate off-line metrology methods to support processes' quality with a \textit{tolerable} assessment of the measurement uncertainty. On the one hand specific-area references propose methods that are not yet well optimized to the changed background, and on the other, international general recommendations guide to effective uncertainty evaluation, but suggesting procedures that are not necessarily proven efficient at the micro- and nano-dimensional scale.
The well-known \textit{GUM approach} (i.e. frequentist statistics) was analyzed with the aim to test consistently its applicability to micro/nano dimensional and surface topography measurements. The investigation assessed three different clarifying situations, giving rise to consistent model equations, and to the achievement of the traceability. The choice of the cases provided a number of influence factors, which are typical liabilities at the micro and nano-length scale, and that have been related to the correction of the systematic behavior, viz. the amount of repeated measurements, the time sequence of the acquired micrographs and the instruments used.
Such approach allowed the successful implementation of the \textit{GUM approach} to micro/nano dimensional and topographic measurements, and also the appraisal of the level of efficacy of the method, its application limits and hints on possible future developments.
\end{abstract}

\pacs{06.20.-f, 06.20.Dk, 02.70.Rr, 02.60.Ed, 02.50.Tt, 02.50.Sk, 07.05.Fb, 07.60.Pb}

\vspace{2pc}

\noindent{\it Keywords}: systematic behavior, frequentist statistics, uncertainty modeling, GUM, surface metrology, optical microscopy, surface topography measuring instruments




\section*{Introduction}\label{intro}
\setcounter{section}{1}
In recent years, the need for shared rules for uncertainty evaluation in manufacturing metrology at the micro and nano dimensional scales has been discussed largely in a comparison of sixteen optical instruments from thirteen international research laboratories measuring the surface topography by areal parameters \cite{comparison}. Among the important issues emphasized by the comparison, it emerged clearly the lack of specific guidelines for the uncertainty evaluation of measurements from optical instruments. It resulted underestimated for the most, commonly based on the repeatability. However, in few cases the final uncertainty was stated with established traceability, but the use of non-suitable artifacts for the sub-micrometer length scale reversed the result towards an overestimated assessment.

The comparison, being the first one on areal texture parameters, also asserted the increasing interest for a three-dimensional (3D) characterization (\textit{areal method}), which offered a more realistic representation of the surfaces, and a more complete understanding of their functions \cite{hansen1}--\cite{Leach-book}. In fact, despite the \textit{profile method} was well established \cite{ISO3274}--\cite{ISO12179} (a new framework for the profile method is currently under development in the new series ISO 21920), it also became insufficient to determine the exact nature of 3D topographic features, and to cope with the constant reduction of critical dimensions, as it is inexorably occurring in micro and nano manufacturing \cite{tosello1}--\cite{tosello3}. Nonetheless, the impact of a non-proper evaluated measurement uncertainty, and the challenge in establishing the traceability were detrimental for the assessment of the manufacturing quality \cite{weckenmann}--\cite{kunzmann}.

Since then the technological rush has prompted into a digital manufacturing era with almost endless possibilities of 'true 3D', complex and multifunctional surfaces \cite{jiang}--\cite{leach1}, and with a new context for the instruments apt to measure the surface topography in the series ISO 25178 based on the metrological characteristics \cite{leach2}--\cite{ISO25178-600}. Even though different working principles exist, the metrological characteristics define sources of variation of the output in common to all the areal topography measuring instruments expressed as uncertainty contributors. Examples of earlier applications can be found elsewhere \cite{giusca1}--\cite{macaulay}, while a new ISO standard is currently under development, where the calibration of the metrological characteristics of areal topography measuring instruments is established through measurements of material measures \cite{ISO25178-700}. Moreover, an additional metrological characteristic has been introduced recently to model the interaction between the surface under measurement and the instrument (topography fidelity---such quantity was commonly referred to in jargon as \textit{surface-instrument convolution}). Eventually, the correction of the systematic effect is achieved as adjustment from the deviations with respect to the measured material measures. The state of the art of the metrological characteristics' framework has been illustrated in a thorough review \cite{leach3}.

The use of metrological characteristics has many advantages. It is a unique framework with clear recommendations that could be adopted easily in any environment. Nonetheless, some ambiguities are still unsolved.

There are no indications regarding the topography fidelity assessment, and what kind of contributor it should be \cite{ISO25178-700}--\cite{leach3}. The correction of the systematic behavior ({\em systematics}) is proposed as adjustment, while it would be laying just in the instrument-surface interaction, which the topography fidelity aims to describe. In addition, the correction appears quite partial, achieved on the known systematic behavior in comparison with material measures. Most important, the whole framework relies on material measures, which are inescapably affected by manufacturing imperfections, and by quick aging and defects depending on the material used for the manufacture (nickel, steel, silicon substrates, etc.). They are also expensive, difficult to find in small dimensions and high aspect ratio above all at the nanoscale, forcing to an extrapolation towards the lower part of the length scale, enlarging even more the undetected variability.
Material measures for use in manufacturing are classified in the geometrical product specification (GPS) system (see ISO~25178-70 \cite{ISO25178-70}), while typical examples and extensive considerations can be found elsewhere \cite{DeChiffreArtefact}--\cite{CarmignatoArtefact}.

The correction of the instruments' systematic behavior subject matter of this paper aims to be in agreement with the international recommendations of the well-known ``Guide to the expression of uncertainty in measurement'' (GUM---also GUM approach) \cite{GUM}, where the finally uncertainty statement is consequence of the correction of the systematics modeled by frequentist statistics. 

The GUM approach in micro and nano metrology has already been considered in past works, mainly for the profile method \cite{krystek}--\cite{haitjema}.

Krystek \cite{krystek} analyzed the case of roughness measurements, and investigated the propagation of the measurement uncertainty when filtering was applied. He also explained that filtered data are always correlated even if the unfiltered ones are not.

Morel \cite{morel} and Haitjema \cite{haitjema} pointed out the impossibility of applying the GUM approach to geometric and roughness measurements providing alternatives based on simulation. Haitjema also defined a type A model for the uncertainty assessment of the height values--- considered correlated---in areal topographical acquisitions. Nonetheless, the final equations resulted unmanageable because of the excessive amount of data.
The correction of the systematics was not performed in any of the cases, but Morel and Haitjema considered the compensation of influence factors in line with the instruments' metrological characteristics (amplification, linearity, residual flatness, noise, spatial resolution and probe size).

Eventually, it is worth to mention Moroni et al. \cite{moroni}, who instead simulated the systematic behavior in 3D microscopy to be accounted in a {task-specific} uncertainty estimation.

A gap between the GUM recommendations and what is considerable achievable in manufacturing metrology appears evident. On the one hand, attempts have been made to apply the GUM approach. On the other, the latest developments in the context of micro and nano metrology---series ISO 25178---does not take into account the correction of the systematic behavior and the definition of a model equation, which are instead distinct recommendations in the GUM.

Henceforth, the proposed paper pursues clarifications, highlighting differences and constrains, and concurrently aiming at expressing the gap between GUM and ISO recommendations.

\subsection{Underlying hypotheses and structure of the paper}\label{hyp}
The work in the paper was carried out with the fundamental hypothesis that the analysis performed is valid on averages of the quantities expected to be measured (the measurands \cite{GUM}). In fact, because of the complexity of both surface topography measuring instruments and ensuing output, an ambiguity may arise on the fashion the measurands were defined.
Specifically, surface topography measuring instruments produce in output a set of height values (related to pixels of an imaging sensor), which altogether are a three-dimensional representation of a portion of a surface \cite{Book-chap10}.
To determine the metrological perspective, averages (arithmetic, integral, etc.) are extracted from the set of height values according to the sought surface characteristics. Therefore, the underlying hypothesis was that averaging the height values retains the systematic behavior.
The validity of such hypothesis is clarified in the discussion (see section~\ref{dis}).
Moreover, being the topic strictly related to metrology tasks, the manufacturing processes of the specimens used are only briefly introduced, and thus referred to dedicated sources for more details.

In order to cope with the aforementioned complexity of the surface topography measuring instruments and their measurements, the topic of correcting the systematic behavior was carefully formulated and narrowed down through a sequence of three case studies. Hence, the paper is structured as follows.

Section~\ref{reference} describes the correction of the systematics carried out within the context of what is diffused routine in the field of metrology for manufacturing. Then, section~\ref{t-seq} analyzes the effect on the systematics' correction when the measurement conditions adhered to what was instead considered good practice in the field. The following section~\ref{charact} summarizes the results of the previous two sections in a typical surface characterization task, and proves the reduction of the final measurement uncertainty in consequence of the systematics' correction. Thus, the sequence of case studies is discussed thoroughly in section~\ref{dis}. Eventually, conclusions are drawn in section~\ref{conc}.

\section{Dimensional and topographic metrology of micro structured tool inserts}\label{reference}
\begin{figure}[t]
\centering
\begin{minipage}[c]{.5\textwidth}
    \centering
    \includegraphics[width=75mm]{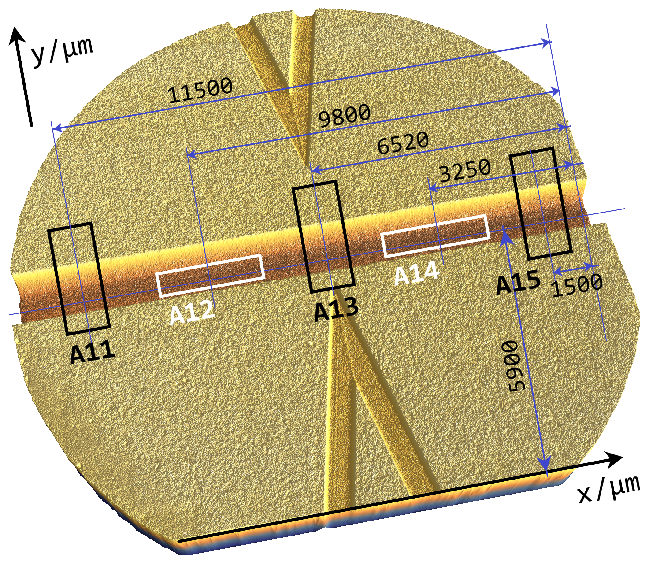}\\
		\vspace{5mm} (a)
  \end{minipage}%
  \begin{minipage}[c]{.5\textwidth}
    \centering
    \includegraphics[width=75mm]{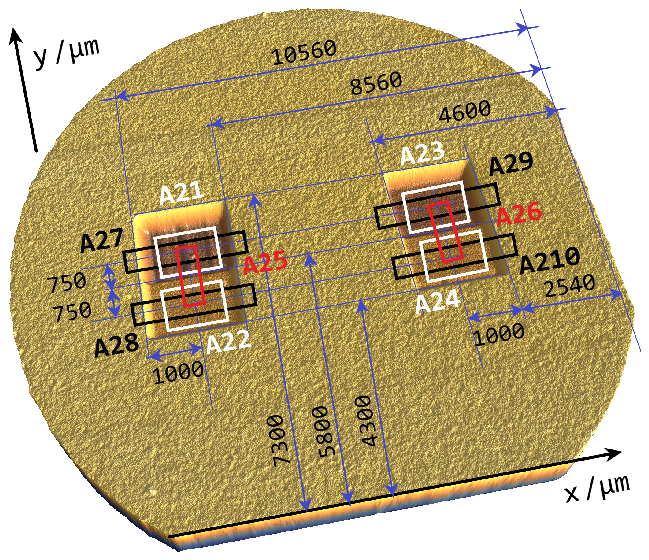}\\
		\vspace{5mm} (b)
  \end{minipage}
  \caption{\label{fig:overview1}Overview of the proof-of-technology micro mold inserts by laser scanning confocal microscope. The regions of areal acquisition were identified over the specimens with respect to a local reference system. Specifically, acquisitions areas were in: (a) A machined straight groove (PoT-1). (b) Two sectioned surfaces at different height (PoT-2).}
	
\end{figure}

Dimensions and micro topographies of proof-of-technology micro mould inserts (PoT) were case in point for an exploratory application of frequentist statistics to measurements from surface topography measuring instruments. The PoT micro mold inserts subject matter within this section are the two steel micro cavities shown in figure~\ref{fig:overview1} as PoT-1 and PoT-2. They have been produced in steel by additive manufacturing and, successively, structured by Jet Electro-chemical Machining (Jet-ECM), in the context of the EU project Hi-Micro (European Commission's 7th Framework Program) \cite{HiMicro}. Details of the manufacturing process can be found in separate papers \cite{hackert1}--\cite{4M-1b}.

The goal in this first instance was to establish a general method for analyzing and, eventually, correcting possible divergences among dimensional and surface topography measurements acquired using different optical instruments. Areal acquisitions from different optical technologies require considering specific critical aspects to perform a correct processing of the measurement data sets, and to allow for an effective comparison among the different instruments.
Such matter has already been emphasized in past works. Mattsson et al. \cite{mattsson} showed that agreement among surface roughness measurements can be limited by both inaccuracies in the definition of the acquisition areas, and the data set assessment, or post-processing. Moreover, Leach et al. \cite{leach4} stated that a comparison among heterogeneous measurements (i.e. measurements from different topography measuring instruments) is only possible when they have the same bandwidth. The bandwidth has been intended in the frequency domain, i.e. the one resulting from a \textsl{Discrete Fourier Transform} (DFT) applied to the set of height values in a micrograph. In the space domain it corresponds to the micrographs' sampling interval. Since the sampling interval may be non-uniform when considering the $x$ and $y$ directions in the micrographs' reference area, henceforth, it will be generically referred as quantization level. Thus, the same quantization level can be achieved practically when the areal acquisitions have (or have been tailored to) the same sampling length and sampling width or, equivalently, the same field of view and same number of height elements per direction.
Yet, the purpose here was to test the eligibility of the approach, inspecting if discrepancies, which are commonly due to the instrument-operator chain (magnification, quantization levels, etc.), could be corrected as systematic differences against a calibrated reference (CI).  In fact, disregarding differences in heterogeneous micrographs is still a quite diffused habit when an exchange of topography measurements happens among laboratories, which have different capabilities for facing the same metrology task. Therefore, in this first implementation, the choice of the magnification (thus, of the quantization level) was left to the single operator.

Measurements of the micro cavities have been carried out inside the straight groove machined on PoT-1 according to the areas indicated in figure~\ref{fig:overview1}-a, namely the height of the steps identified as \textsl{A11}, \textsl{A13} and \textsl{A15}, and the surface texture indicated as \textsl{A12} and \textsl{A14}. While the sectioned surfaces on PoT-2 shown in figure~\ref{fig:overview1}-b comprised the measurand areas \textsl{A21}--\textsl{A24} (surface texture) and \textsl{A25}--\textsl{A210} (height of step). The regions of areal acquisition were identified specifically by defining the on-specimen reference systems shown in figure~\ref{fig:overview1}. The height of the steps was extracted from the micrographs by the histogram technique, namely as distance between the two planes containing the maximum of sampled point occurrences at the corresponding height levels \cite{Book-chap10}. While the surface texture measurands were evaluated averaging the height values on the corresponding micrographs as root mean square value (RMS---$Sq$ areal parameter according to ISO~25178-2 \cite{ISO25178-2}).

Different sessions were performed in three different laboratories using respectively focus-variation microscopy (FV), laser scanning confocal microscopy (LSC) and coherent scanning interferometer microscopy (CSI). In addition, contact measurements by a calibrated stylus instrument (CI) were used as reference for indirectly achieving the traceability. Thus, the measurement uncertainty was garnered from the correction of the systematic behavior, as the discrepancy between the optical instruments' measurements and the calibrated contact ones (rough data are in \ref{appA}).

The  data sets were assessed using the same commercial software \cite{SPIP}, dealing with plane correction, presence of noise and non-measured pixels, influence of filtering, waviness, which are all described in depth elsewhere \cite{4M-1b}. Moreover, no filtering was applied. The application of filters to topography measurements is discouraged here \cite{Book-chap10}, especially concurrently with the application of the frequentist approach, because it would reshape the experimental distribution and, therefore, hinder the statistical analysis. An exception is the form correction, when needed, which is a merely \textit{non-destructive} translation of pixels in the micrographs. Nonetheless, the form correction was not required here. An outlier-removal criterion was instead applied to the extracted averaged measurands. The presence of outliers in the experimental distribution would counteract the efficacy of the normality tests. Barbato et al. suggested divers outlier removal methods in a comprehensive work \cite{barbato}. Who is writing deemed the Chauvenet's criterion more efficient with surface topography areal averages. No more than two iterations of the criterion were fulfilled.

\subsection{Analytical and statistical modeling}\label{statmodel}
The most favorable situation for the evaluation of the measurement uncertainty is a normal distribution of the experimental data, which means that the data are randomly distributed \cite{bailey}. Thus, the frequentist inference approach is based on the validity of the Central Limit Theorem (TCL) \cite{GUM}. A large number of replications of a measured quantity (i.e. random variables), which are independent in a sampling distribution (viz. experimental distribution), tends as a whole to a normal one---even if each single random variable is not normally distributed. Conversely, a large number of random variables, altogether rejecting the normality, reveals that a number of them is not independent. Thus, a mathematical relationship can model the dependency among the random variables, which corresponds to the sought systematic behavior inside an experimental distribution of measured quantities. Such mathematical relationship subtracted from the experimental distribution allows to minimize the systematic behavior, and the residual random variables are expected to be statistically quasi-independent, i.e. approaching a normal distribution (it is hardly possible to model and correct the systematic factors completely) \cite{Barbato-book}--\cite{Pavese-book}. The large number of independent random variables (infinite in the mathematical formulation of the theorem) that allows for an approximated nonetheless effective application of the TCL can essentially be found comparing the $t$~distribution---used for small size samples---and the normal one at the same confidence level of 95~\%---commonly used in metrology. The two distributions show the same trend with acceptable approximation when the $t$~distribution has degrees of freedom (DoF) from 10--15 and above. Hence, as a rule of thumb, such inferential statistics should be applied to experimental distributions with more that 10--15 repeated measurements per measurand.

\begin{figure}[ht]
\centering
\begin{minipage}[c]{.5\textwidth}
    \centering
    \includegraphics[width=75mm]{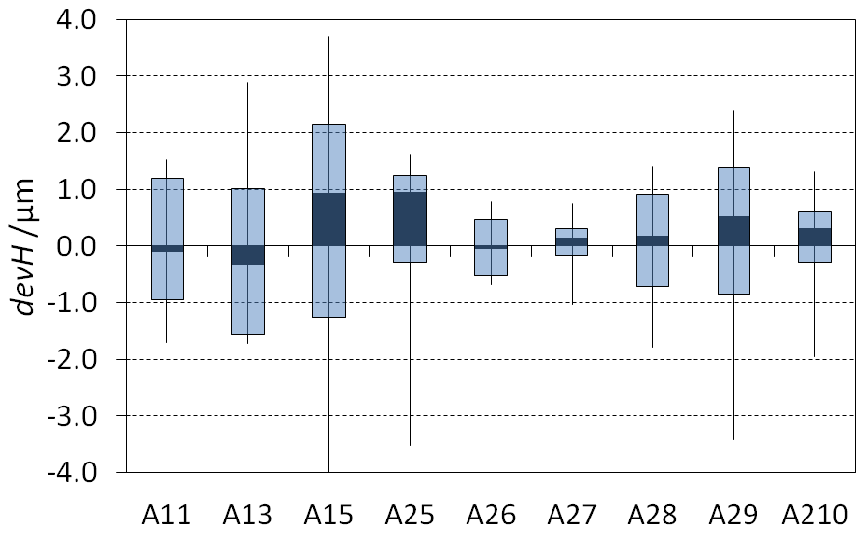}\\
		\vspace{5mm} (a)
    \includegraphics[width=75mm]{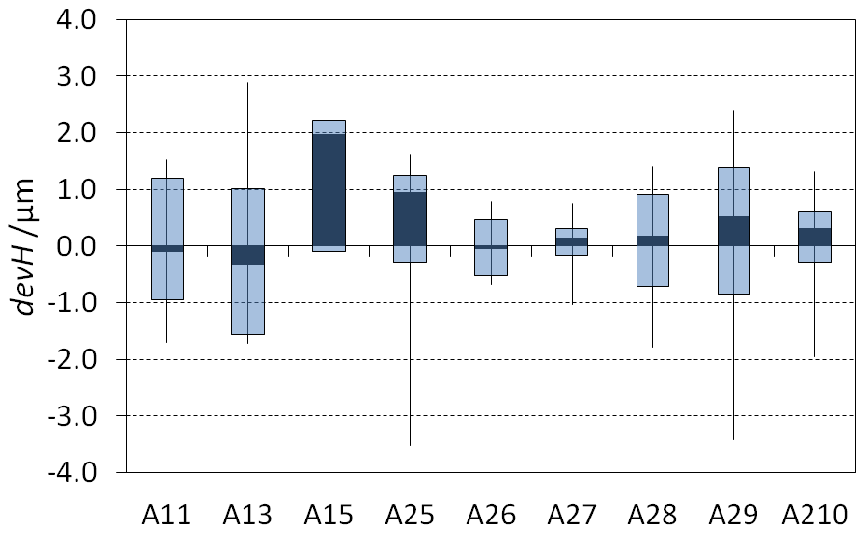}\\
		\vspace{5mm} (c)
  \end{minipage}%
  \begin{minipage}[c]{.5\textwidth}
    \centering
    \includegraphics[width=75mm]{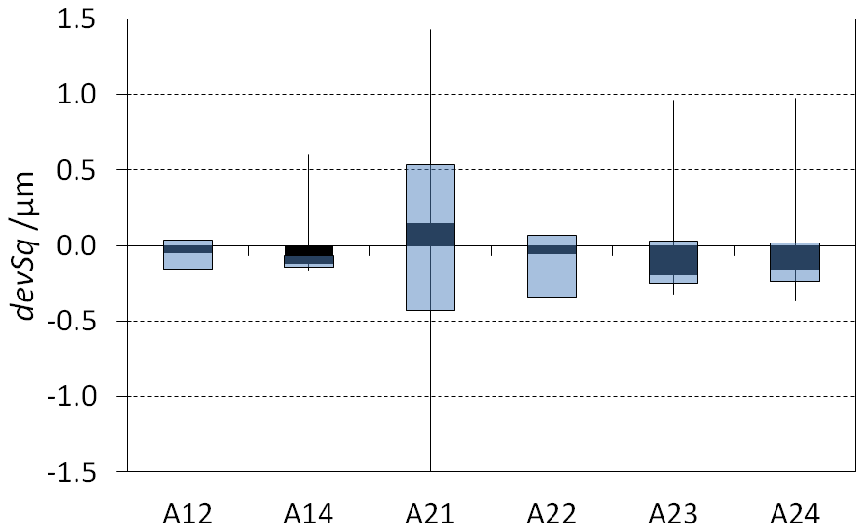}\\
		\vspace{5mm} (b)
    \includegraphics[width=75mm]{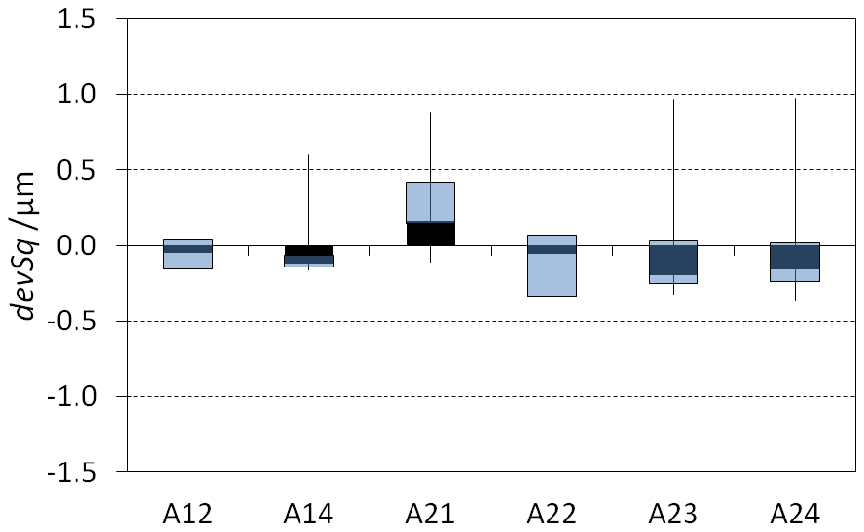}\\
		\vspace{5mm} (d)
  \end{minipage}
	\caption{\label{fig:boxplot1}Box-plots of deviations with inter-quartile range (box), maximum (upper whisker), minimum (lower whisker) and median (column in the box) values. (a) Deviations $devH$ of step height before outliers removal. (b) Deviations $devSq$ of root mean square value before outliers removal. (c) Deviations $devH$ of step height after outliers removal (Chauvenet's exclusion limits $x_{L~min}=-3.8{\rm~\mu m}$, $x_{L~max}=4.1{\rm~\mu m}$). (d) Deviations $devSq$ of root mean square value after outliers removal (Chauvenet's exclusion limits $x_{L~min}=-0.9{\rm~\mu m}$, $x_{L~max}=1.0{\rm~\mu m}$).}
\end{figure}

Regarding the PoT, even though the measurement data were a substantial number when considering the areal acquisitions over the specimens, they were limited in terms of repeated measurements and unbalanced, i.e. discordant in terms of replicated measurements (see the raw data after the assessment of the data sets (post-processing) in \ref{appA}, tables~\ref{tab:A:height} and \ref{tab:A:sq}). To obtain a consistent number of data, and in agreement with the starting hypothesis of correcting the systematics for heterogeneous measurements, the results of the post-processing were normalized to their respective areal averages (subtraction of the corresponding averages in \textsl{A11},...,\textsl{A210}, respectively). The normalized deviations are in \ref{appA}, tables~\ref{tab:A:devH} and \ref{tab:A:devSq}. They are also pictured in the box-plots in figure~\ref{fig:boxplot1}-a~and~-b in terms of inter-quartile range (IQR) of the two groups of measurands.

\begin{figure}[ht]
\centering
\begin{minipage}[c]{.5\textwidth}
    \centering
\includegraphics[width=77mm]{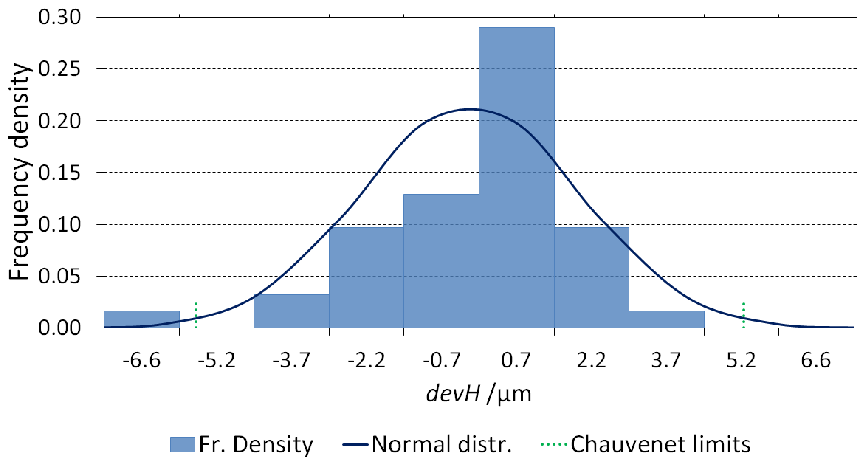}\\
		\vspace{5mm} (a)
\includegraphics[width=77mm]{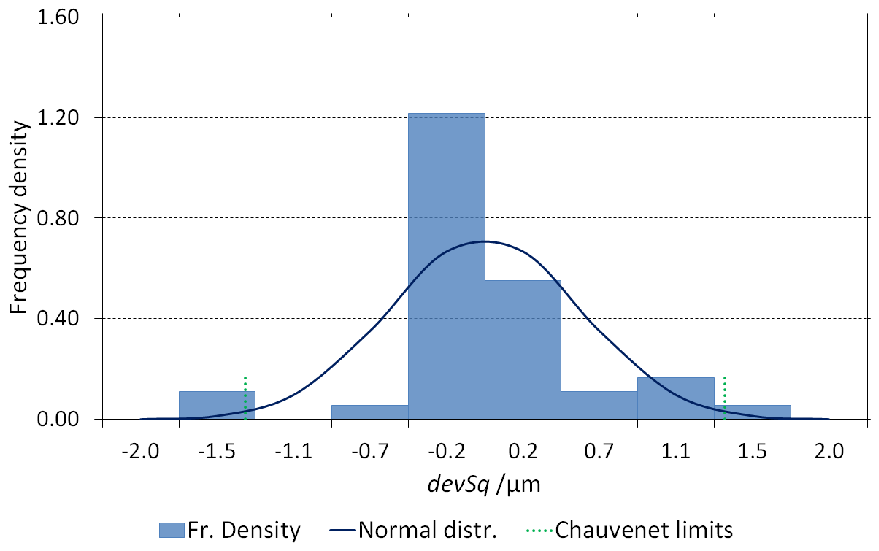}\\
		\vspace{5mm} (c)
  \end{minipage}%
  \begin{minipage}[c]{.5\textwidth}
    \centering
\includegraphics[width=77mm]{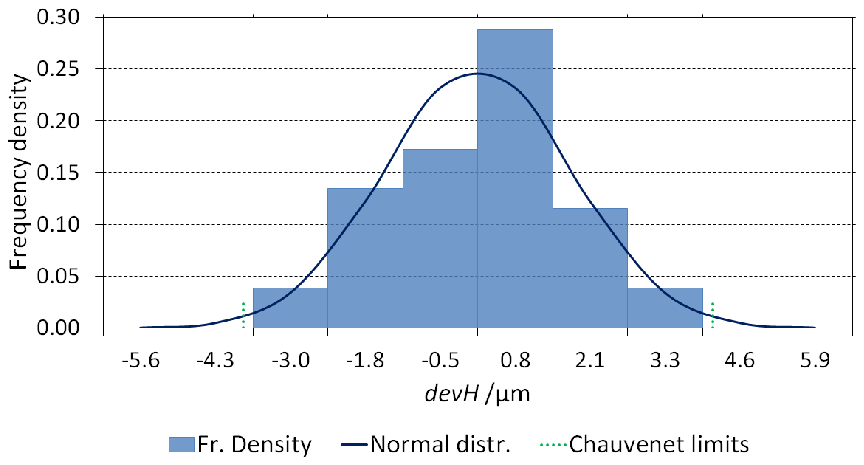}\\
		\vspace{5mm} (b)
\includegraphics[width=77mm]{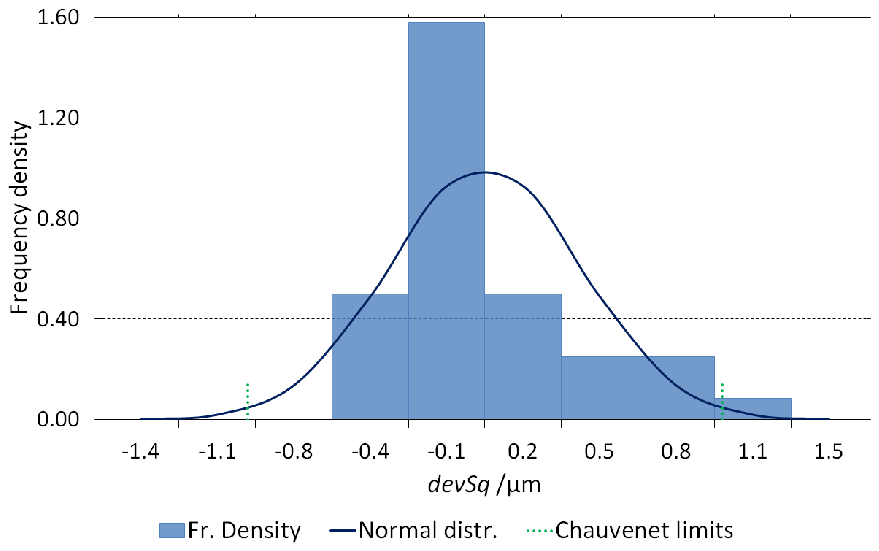}\\
		\vspace{5mm} (d)
  \end{minipage}
	\caption{\label{fig:hist1}Histograms of the experimental distribution, and comparable curves of the theoretical normal distribution for the deviations $devH$ of step height---before (a) and after (b) the outliers removal (Chauvenet's exclusion limits $x_{L~min}=-3.8{\rm~\mu m}$, $x_{L~max}=4.1{\rm~\mu m}$)---and for the deviations $devSq$ of root mean square value---before (c) and after (d) the outliers removal (Chauvenet's exclusion limits $x_{L~min}=-0.9{\rm~\mu m}$, $x_{L~max}=1.0{\rm~\mu m}$).}
\end{figure}

The exclusion principle was applied next. In order to avoid the risk of excluding also suitable results, a full understanding of the experimental data was attained monitoring box-plots and histograms for, respectively, inspecting agreement among the experimental data and visualizing their distribution (see figures~\ref{fig:boxplot1}~and~\ref{fig:hist1}). The elimination was confirmed by the presence of disturbances in corresponding micrographs. The values excluded by two iterations of the Chauvenet's criterion were four, and are indicated in parenthesis in the mentioned tables of \ref{appA}.
After the exclusion, the deviations were 41 representing the step heights and 38 the surface texture, and almost all the IQRs were in good agreement in each respective group, which means that the exclusion was effective. An exception was the area \textsl{A21}. A closer look to the surface (see figure~\ref{fig:overview2}-b in the next section~\ref{t-seq}) revealed some waviness only present in this measurand area. Nonetheless, no further value was evidenced by the exclusion principle. Moreover, no reasons were found to justify a third iteration of the Chauvenet's criterion.

\begin{figure}[ht]
\centering
\begin{minipage}[c]{.5\textwidth}
    \centering
    \includegraphics[width=75mm]{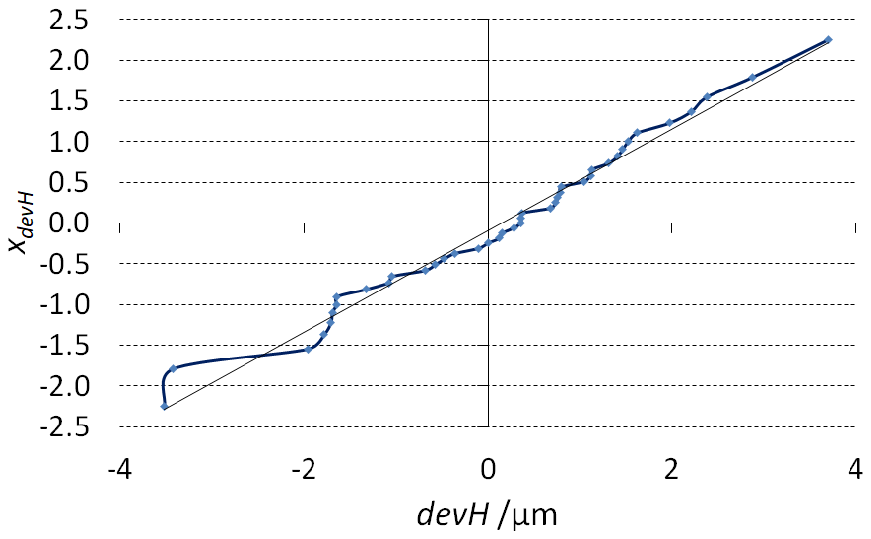}\\
		\vspace{5mm} (a)
  \end{minipage}%
  \begin{minipage}[c]{.5\textwidth}
    \centering
    \includegraphics[width=75mm]{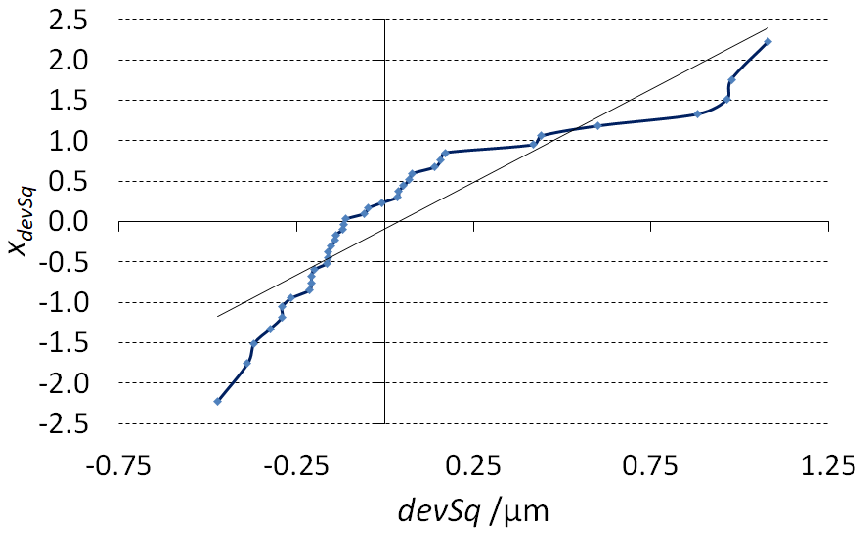}\\
		\vspace{5mm} (b)
  \end{minipage}
  \caption{\label{fig:NPP}Normal Probability Plot (NPP). (a) NPP of deviations of step height $devH$. (b) NPP of deviations of root mean square value $devSq$.}
\end{figure}   

\begin{table}[ht]
\caption{\label{tab:ANOVA}General linear model ANOVA of deviations $devH$ and $devSq$ (adjusted sum of squares---confidence level 95~\%).}
\footnotesize
\begin{indented}
\lineup
\item[]\begin{tabular}{@{}lll}
\br
							&$devH$																&$devSq$\\
\mr
Instrument		& Influence	(p-val $< 0.01$)          & No influence (p-val $\approx 0.16$)\\
Magnification	& Influence (p-val $\approx 0.03$)    & Influence (p-val $< 0.01$)\\
Area					& No influence (p-val $\approx 0.69$) & No influence (p-val $\approx 0.25$)\\
$R^{2}$				& 61~\%                               & 84~\%\\
\br
\end{tabular}\\
\end{indented}
\end{table}
\normalsize

After the outliers removal, a departure of the experimental distributions from the comparable normal ones (theoretical normal distributions with same average and standard deviation of the corresponding experimental ones) can be deducted by inspecting the histograms in figure~\ref{fig:hist1}-b~and~-d. In addition to the histograms, the normal probability plots (NPP) have trend clearly far from a straight line (see figure~\ref{fig:NPP}). Looking at the extremes of the curves, they show different trends of slope in both diagrams. Moreover, the slopes are not confirmed in the middle, where an arc is present. It is more pronounced and concave downwards for the deviations $devSq$, which evidences a right asymmetric distribution (right-skewed). The arc is instead almost unnoticeable with opposite orientation for the deviations $devH$, confirming a left-skewed distribution.
Nonetheless, it is difficult to analyze completely the NPP curves because of their complexity, which suggests that different slopes piecewise can be recognized (multimodal distributions), thus presuming that the systematic behavior was due to several factors.

An analysis of variance (ANOVA) was performed to relate the systematic behavior to the factors involved in the measurements. Being the data sets unbalanced, a general linear model was used to implement the ANOVA test, considering the three factors: \textit{instrument}, \textit{magnification} and \textit{measurand area}. The test was performed using adjusted sum of squares, which means that the test was not considered dependent on any order of the factors into the model (the laboratories did not provide information regarding the order of acquisition), and is summarized in table~\ref{tab:ANOVA}. Moreover, the experimental data were not enough for also considering the factors' interactions. For the sake of completeness, the ANOVA test was also performed using sequential sum of square, in which case the \textit{instrument} was an influence factor for $devSq$, too (p-val $\approx 0.02$). Since no information was available about the sequence in which the data were acquired, the given order was an arbitrary one, nonetheless quite suitable to show a possible incompleteness of the test. In addition, the coefficients of determination $R^{2}$ associated to the ANOVA models were not very robust, thus the models were not well fitted.

\begin{figure}[ht]
\centering
\begin{minipage}[c]{.5\textwidth}
    \centering
    \includegraphics[width=75mm]{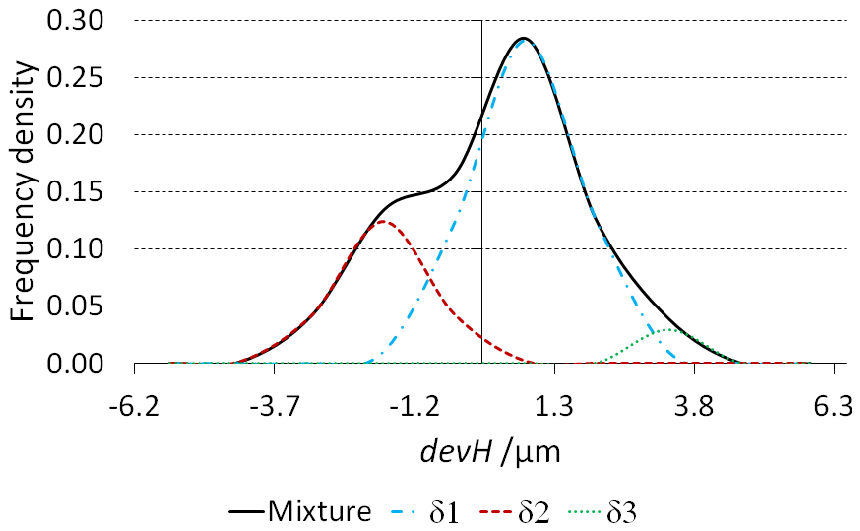}\\
		\vspace{5mm} (a)
    \includegraphics[width=75mm]{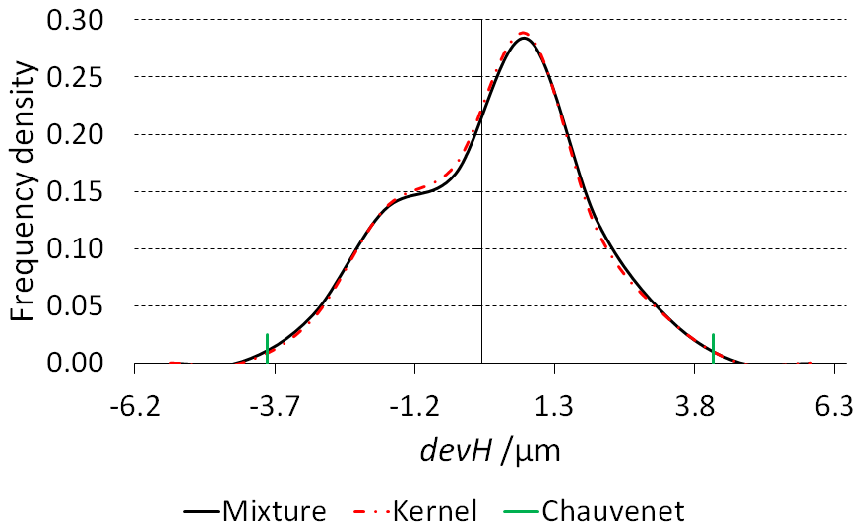}\\
		\vspace{5mm} (c)
  \end{minipage}%
  \begin{minipage}[c]{.5\textwidth}
    \centering
    \includegraphics[width=75mm]{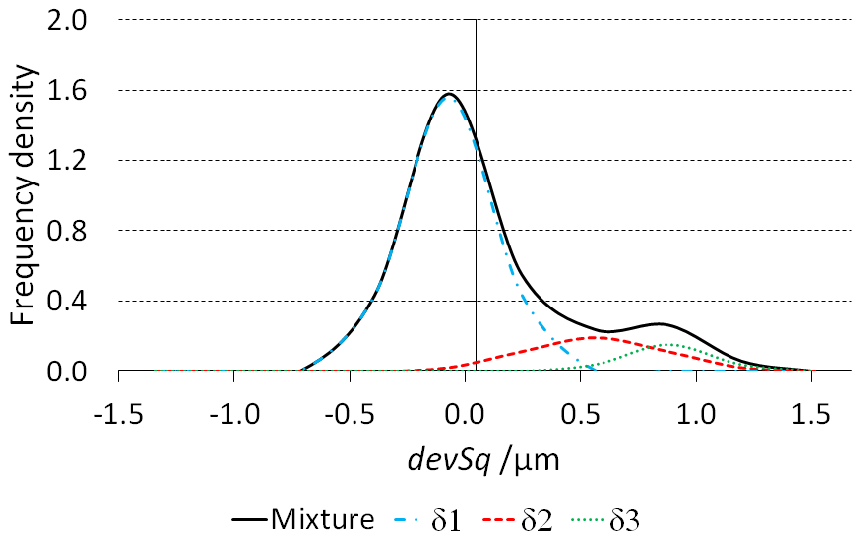}\\
		\vspace{5mm} (b)
    \includegraphics[width=75mm]{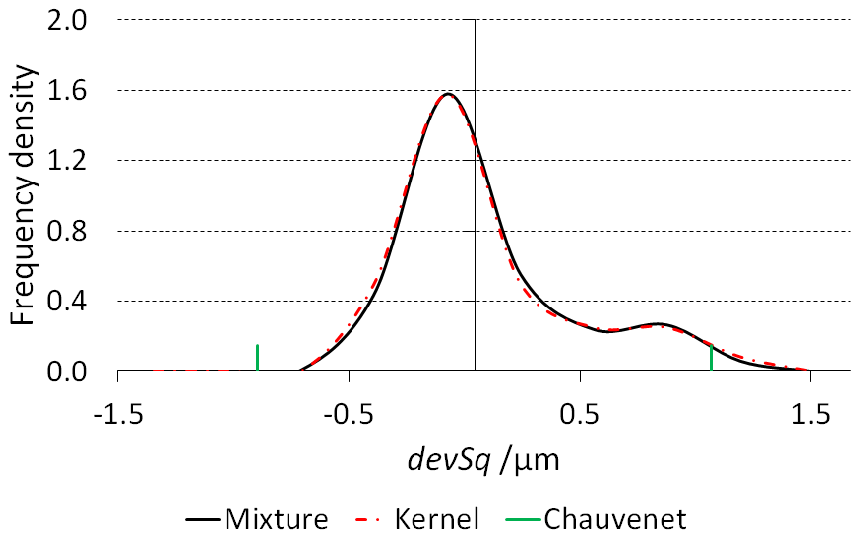}\\
		\vspace{5mm} (d)
  \end{minipage}
  \caption{Comparison between mixture and estimation of the probability density of the experimental data (kernel distribution). (a) Mixture of deviations of step height $devH$. (b) Mixture of deviations of root mean square value $devSq$. (c) Comparison between kernel distribution and kernel density estimation (mixture) of deviations of step height $devH$. (d) Comparison between kernel distribution and kernel density estimation (mixture) of deviations of root mean square value $devSq$.}
	\label{fig:mixture1}
\end{figure}   

\begin{table}[ht]
\caption{\label{tab:mixture:devH}Parameters of the individual kernels in the mixture of deviations $devH{\rm~/\mu m}$ (the optimised $\chi^2$ statistics is 0.55).}
\footnotesize
\begin{indented}
\lineup
\item[]\begin{tabular}{@{}llll}
\br
Individual ker.&Average${\rm~/\mu m}$&St. Dev.${\rm~/\mu m}$&Percent. Of incid.~/\%\\
\mr
$\delta_1$ & $0.79$ & $0.99$ & $70$\\
$\delta_2$ & $\-1.76$ & $0.87$ & $27$\\
$\delta_3$ & $3.33$ & $0.45$ & $3$\\
\br
\end{tabular}\\
\end{indented}
\end{table}
\normalsize

\begin{table}[ht]
\caption{\label{tab:mixture:devSq}Parameters of the individual kernels in the mixture of deviations $devSq{\rm~/\mu m}$ (the optimised $\chi^2$ statistics is 0.77).}
\footnotesize
\begin{indented}
\lineup
\item[]\begin{tabular}{@{}llll}
\br
Individual ker.&Average${\rm~/\mu m}$&St. Dev.${\rm~/\mu m}$&Percent. Of incid.~/\%\\
\mr
$\delta_1$ & $\-0.12$ & $0.20$ & $78$\\
$\delta_2$ & $0.51$ & $0.31$ & $15$\\
$\delta_3$ & $0.83$ & $0.19$ & $7$\\
\br
\end{tabular}\\
\end{indented}
\end{table}
\normalsize

To solve this indeterminacy, the joint effect of random and systematic factors was described as a mixture of normal distributions \cite{aggogeri}. Therefore, a deeper analysis was performed decomposing the envelope of a histogram (kernel distribution), which represent the frequency density of the experimental data (probability density estimation of a measurand), into a mixture of theoretical normal distributions (individual kernels). The decomposition (kernel density estimation---KDE) was obtained by an optimization of the $\chi^2$ test (minimization of $\chi^2$ statistics) performed on the two measurands' deviations statistically grouped in histograms.
Figure~\ref{fig:mixture1} shows the mixtures of three individual kernels for each measurand's deviations, and the comparison between the estimated mixtures and the kernel distributions. As shown in the figure, the mixtures describe the kernel distributions' shape with a satisfactory fidelity. This suggested the presence of three dominant influence factors, even though the presence of other unknown systematic factors could not be excluded \cite{bailey}--\cite{Pavese-book}. The estimate of the parameters of the individual kernels in the mixtures (average, standard deviation and percentage of incidence) is given in the tables~\ref{tab:mixture:devH}~and~\ref{tab:mixture:devSq}.

\subsection{Modeling the systematic behavior}\label{syst}
\begin{figure}[ht]
\centering
\begin{minipage}[c]{.5\textwidth}
    \centering
\includegraphics[width=77mm]{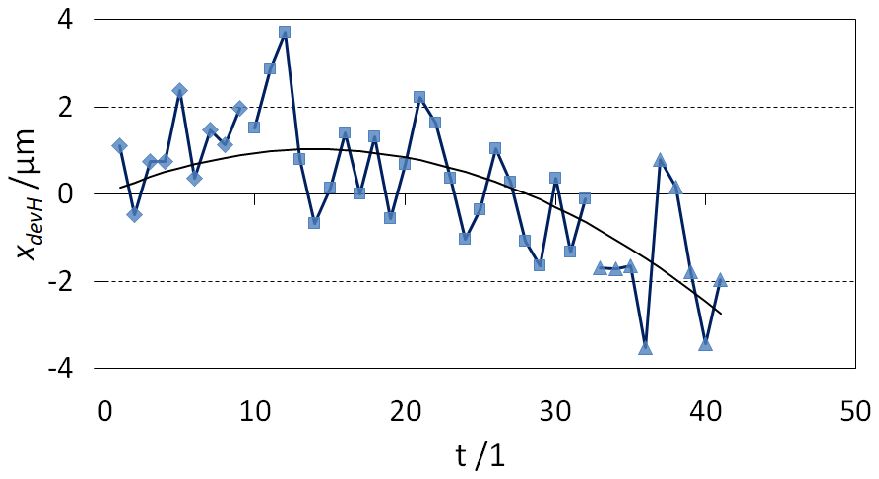}\\
		\vspace{5mm} (a)
  \end{minipage}%
  \begin{minipage}[c]{.5\textwidth}
    \centering
\includegraphics[width=77mm]{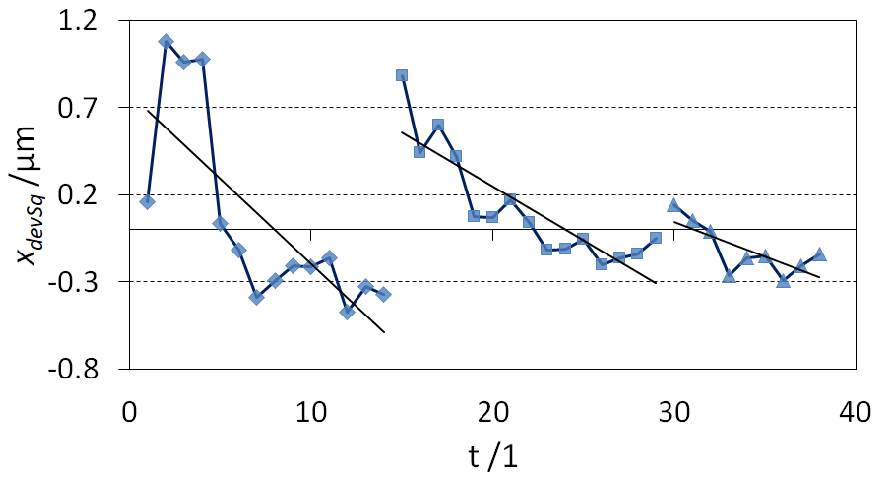}\\
		\vspace{5mm} (b)
  \end{minipage}
	\caption{\label{fig:sequence}Diagram of the random variable $x_{devH}$ of the deviations of step height (a), and the one of the random variable $x_{devSq}$ of the deviations of root mean square value (b), both represented as function of the arbitrary sequences chosen. On the left of each diagram there is the sequence \fulldiamond~related to LSC; in the middle the one \fullsquare~related to CSI; on the right the sequence \fulltriangle~of FVM.}
\end{figure}

Figure~\ref{fig:sequence} shows the deviations $devH$ and $devSq$ represented according to the sequence arbitrarily chosen for the current investigation. Their distributions clearly show tendencies that could be identified by regression models, and successively corrected (see next section~\ref{t-seq}). Changing the sequence of the data the tendencies also change. This can affect the results of an ANOVA test but not the correction of the systematic behavior. In fact, the particular mathematical model used for correcting a sequence of repeated measurements is not influential for the correction if the data are always referred to that specific model equation. In addition, in the procedure proposed in this section, the correction was achieved against the reference measurements CI, which was intended accounting for any detectable systematic behavior.

Thus, a least squares regression was implemented onto the non-normalized values of deviations (de-normalized by their respective areal averages) to evaluate discrepancies between optical and contact measurements, i.e. the reference CI.
The model equation found consistent with the experimental data of both step height $H$ and RMS $Sq$ was a straight line passing by the origin. In fact, if different instruments measure the same quantity, ideally, the results should equal (unitary angular coefficient). Hence, the best fit evaluated the mismatch between $x_{OPT}$, indicating any optical instrument measurements, and $x_{CI}$, i.e. the reference, with respect to the unitary slope

\begin{equation}
x_{OPT} = q \cdot x_{CI}
\label{mat_eq}
\end{equation}

Equation~(\ref{mat_eq}) was found consistent with the experimental data to mathematically describe the best-fit regression. Other factors influencing the evaluation were the reproducibility $\epsilon_{Rep}$, i.e. the standard deviation of the regression results, and the resolution $\epsilon_{res}$. Thus, the metrological model equation used for the uncertainty evaluation was

\begin{equation}
y_{OPT} = q \cdot x_{CI} \pm \epsilon_{Rep} \pm \epsilon_{res}
\label{met_eq}
\end{equation}

where reproducibility and resolution were considered random variables with null average.

The instrument resolution is usually included in the reproducibility through repeated measurements (unless a coarse increment in basic instruments requires to be treated separately).
Nevertheless, separate values for reproducibility and resolution may be considered when they refer to different contributors in the metrological chain, which is actually the case considered in this section. In fact, despite the stylus instrument resolution could be considered non-influential, the raw CI measurements were also analyzed by a post-processing software \cite{SPIP}. For this reason, the resolution was set equal to the numerical precision of the post-processing software. This was a conservative choice because, even though unlikely, the software significant figures might limit the instrument resolution.

Hence, the standard uncertainty was calculated propagating the uncertainty contributors to (\ref{met_eq}) by the usual method of combination of variances for uncorrelated quantities, where the contributors were as follows:

\begin{itemize}
\item	The accuracy of CI ($U=0.112\rm~\mu m$, unfiltered) stated in the calibration certificate of the instrument for the range of interest (cf. \ref{appC}, too).
\item	The standard deviation of the coefficient of the model equation assessed in the best fit regression (see $\sigma_q$ in table~\ref{tab:regression}).
\item	The reproducibility of the regression (see table~\ref{tab:regression}).
\item	The post-processing software numerical precision (estimated to be 1~nm---conservative choice).
\end{itemize}

Finally, the expanded uncertainty was evaluated as the confidence interval corresponding to the conventional confidence level of 95~\%, and by a coverage factor calculated using the $t$-distribution with degrees of freedom given by the Welch-Satterthwaite formula \cite{GUM}.

\subsection{Results of the correction}\label{corres}
\begin{figure}[t]
\centering
\begin{minipage}[c]{.5\textwidth}
    \centering
\includegraphics[width=77mm]{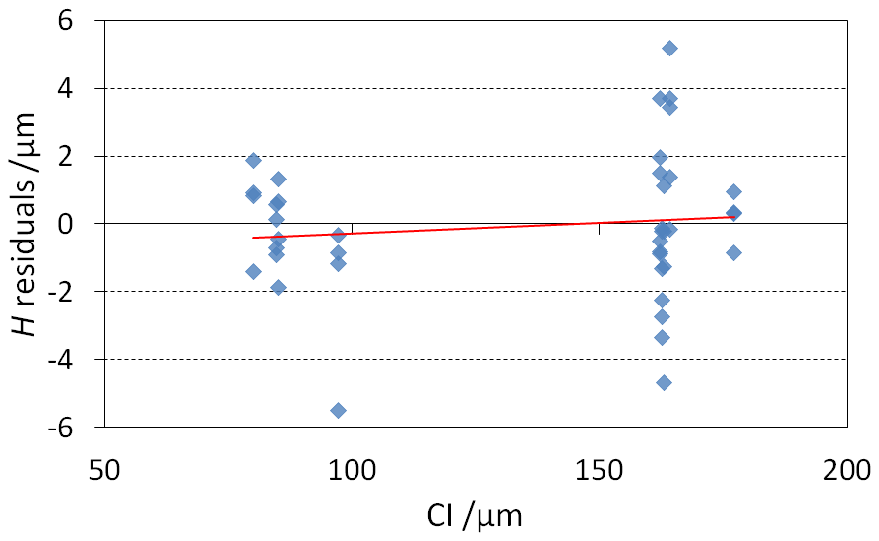}\\
		\vspace{5mm} (a)
  \end{minipage}%
  \begin{minipage}[c]{.5\textwidth}
    \centering
\includegraphics[width=77mm]{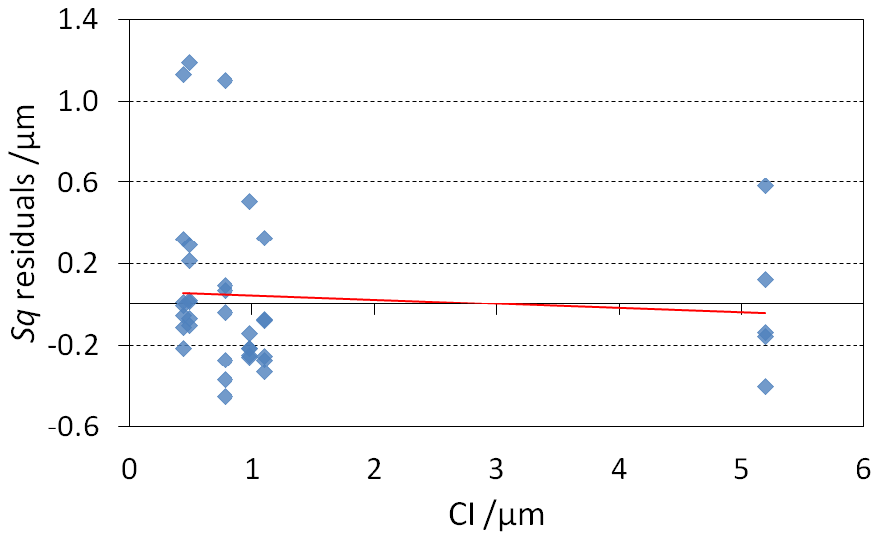}\\
		\vspace{5mm} (b)
  \end{minipage}
	\caption{\label{fig:regression}Residuals $x_{devH}$ of the deviations of step height (a), and the one of the random variable $x_{devSq}$ of the deviations of root mean square value (b), both represented as function of the arbitrary sequences chosen. They are function of the CI. The straight line indicate the trend (first-order polynomial).}
\end{figure}

The results of the correction using the regression models are summarized in the tables~\ref{tab:H:res}~and~\ref{tab:Sq:res}, with values of the reference CI, of the optical instruments and of the evaluated expanded uncertainties. The residuals of the correction are in figure~\ref{fig:regression}. As revealed by the residuals' trend, the correction was not very effective, achieving in both cases a partial compensation of the systematic behavior. Therefore, for highlighting the efficiency of each instrument-operator chain an uncertainty was also evaluated considering a separate regression model for each optical instrument. Such uncertainties were evaluated similarly to the expanded uncertainty, nonetheless, they just had the intent to substantiate the performance of each single instrument-operator chain. It was found $U_{LSC}\approx 3\rm~\mu m$, $U_{CSI}\approx 4\rm~\mu m$ and $U_{FV}\approx 6\rm~\mu m$ for step height measurements (maximum values). While it was $U_{LSC}\approx 1.7\rm~\mu m$, $U_{CSI}\approx 0.7\rm~\mu m$ and $U_{FV}\approx 0.6\rm~\mu m$ for RMS measurements (maximum values). Thus, the \textsl{closest} instrument to the reference CI was LSC for step height measurements, and FV for RMS measurements.

\begin{table}[ht]
\caption{\label{tab:regression}Parameters of the least square regressions---first order. Slope $q$, standard deviation of the slope $\sigma_q$, reproducibility $\epsilon_{Rep}$ (standard deviation of the residuals), degrees of freedom and coefficient of determination $R^2$.}
\footnotesize
\begin{indented}
\lineup
\item[]\begin{tabular}{@{}llllll}
\br
          & {$q$~/1} & {$\sigma_q$~/1} & {$\epsilon_{Rep}{\rm~/\mu m}$} & {DoF} & {$R^2$~/\%}\\
\mr
$y_{OPT,H}$ & $0.998$ & $0.002$ & $2.092$ & $40$ & $99.9$\\
$y_{OPT,Sq}$ & $1.038$ & $0.033$ & $0.405$ & $37$ & $88.8$\\
\br
\end{tabular}\\
\end{indented}
\end{table}
\normalsize

\begin{table}[ht]
\caption{\label{tab:H:res}Results of the step height measurements $H{\rm~/\mu m}$ for each measurand area, and related expanded uncertainties. The decimals in parentheses explicate the approximation on the last digit.}
\footnotesize
\begin{indented}
\lineup
\item[]\begin{tabular}{@{}lllllllll}
\br
       & $CI$    & $LSC 5\times$ & $LSC 10\times$ & $CSI 10\times$ & $CSI 20\times$ & $CSI 50\times$ & $FV 20\times$ & $U$\\
\mr
$A11$  & $162.6$ &         & $162.1$  & $162.1$   & $160.0$   & $160.2$     & $158.9$  & $4.3(0)$\\
$A13$  & $162.3$ &         & $163.9$  & $165.7$   & $163.5$   & $161.3$     & $161.1$  & $4.3(0)$\\
$A15$  & $164.2$ &         & $167.3$  & $169.0$   & $167.6$   & $165.2$     & $163.7$  & $4.3(0)$\\
$A25$  & $97.3$  & $96.2$  &          & $95.9$    & $96.7$    &             & $91.6$   & $4.2(6)$\\
$A26$  & $84.8$  & $83.9$  &          & $83.7$    & $84.8$    &             & $85.2$   & $4.2(6)$\\
$A27$  & $177.2$ & $177.8$ &          & $177.1$   & $176.0$   &             & $177.2$  & $4.3(1)$\\
$A28$  & $85.2$  & $85.6$  &          & $86.3$    & $84.5$    &             & $83.1$   & $4.2(6)$\\
$A29$  & $163.1$ & $163.9$ &          & $161.5$   & $162.6$   &             & $158.1$  & $4.3(0)$\\
$A210$ & $80.2$  & $80.9$  &          & $81.9$    & $80.8$    &             & $78.6$   & $4.2(6)$\\
\br
\end{tabular}\\
\end{indented}
\end{table}
\normalsize

\begin{table}[ht]
\caption{\label{tab:Sq:res}Results of the RMS measurements $Sq{\rm~/\mu m}$ for each measurand area, and related expanded uncertainties.}
\centering
\footnotesize
\lineup
\begin{tabular}{@{}lllllllllll}
\br
       & $CI$    & $LSC 5\times$ & $LSC 50\times$ & $LSC 100\times$ & $CSI 10\times$  & $CSI 20\times$ & $CSI 50\times$ & $FV 20\times$ & $FV 50\times$ & $U$\\
\mr
$A12$ & $1.10$ &        & $1.06$ & $0.81$ &        & $1.47$ & $0.98$ &        & $0.87$ & $0.90$\\
$A14$ & $0.98$ &        & $0.80$ & $0.76$ &        & $1.52$ & $0.84$ &        & $0.77$ & $0.90$\\
$A21$ & $5.19$ & $5.25$ &        &        & $5.97$ & $5.51$ & $4.98$ & $5.23$ &        & $0.95$\\
$A22$ & $0.78$ & $1.91$ & $0.44$ & $0.36$ &        & $0.91$ & $0.77$ & $0.88$ & $0.54$ & $0.90$\\
$A23$ & $0.49$ & $1.69$ & $0.44$ & $0.41$ &        & $0.80$ & $0.53$ & $0.72$ & $0.52$ & $0.89$\\
$A24$ & $0.44$ & $1.58$ & $0.40$ & $0.24$ &        & $0.78$ & $0.45$ & $0.34$ & $0.47$ & $0.89$\\
\br
\end{tabular}\\
\end{table}
\normalsize

\section{Correction of the systematic behavior vs time sequence}\label{t-seq}

\begin{figure}[ht]
\centering
\begin{minipage}[c]{.5\textwidth}
    \centering
    \includegraphics[width=60mm]{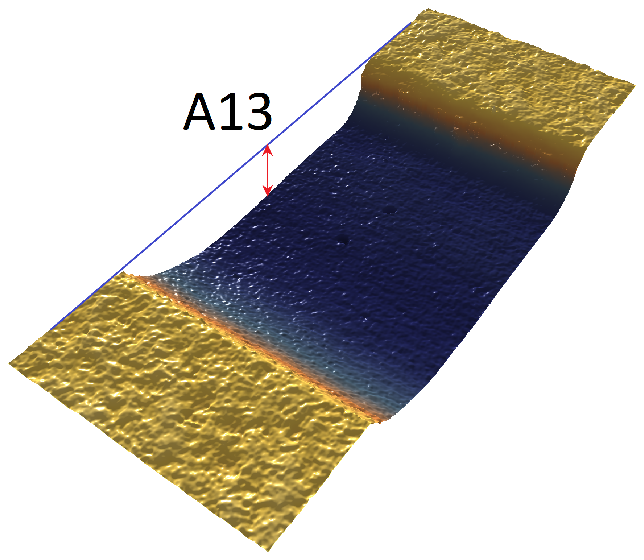}\\
		\vspace{5mm} (a)
  \end{minipage}%
  \begin{minipage}[c]{.5\textwidth}
    \centering
    \includegraphics[width=90mm]{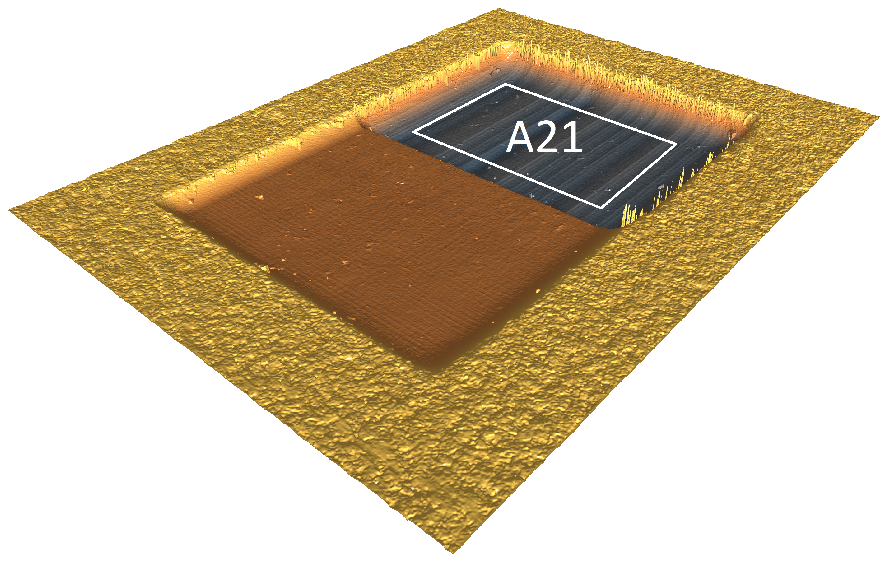}\\
		\vspace{5mm} (b)
  \end{minipage}
  \caption{\label{fig:overview2}Close-up of the measurand areas \textsl{A13} (a) and \textsl{A21} (b).}
\end{figure}

In the previous instance, because of a shortfall of measurements, the correction of the systematic behavior and the establishment of the traceability were only possible against a reference (calibrated/\textit{calibratable} instrument), matching the correspondent measurements. Nevertheless, if reference and instrument under test show incompatible trend the correction becomes inadequate (or impossible, as in the previous case).

Indeed, the development of a systematic behavior is to be sought in the non-ideal behavior and use of a measurement instrument, in actual environmental conditions, and independently from any reference. Hence, such behavior is expected developing during the time of use, building up at each occurrence of the measurement events and, thus, is strictly related to their acquisition sequence.

To better inspect this matter, the measurand areas \textsl{A13} (height of step---figure~\ref{fig:overview1}-a) and \textsl{A21} (RMS value of texture---figure~\ref{fig:overview1}-b) of the same specimens in the previous study were re-measured and corrected for the systematics according to the time sequence of acquisition (see also close-up of those measurand areas in figure~\ref{fig:overview2}).
Several repeated measurements were performed by FV using different magnifications ($5\times$, $10\times$, $20\times$, $50\times$ and $100\times$).  Twenty-five repeated areal acquisitions were in the measurand area \textsl{A13}, and thirty in \textsl{A21}. Rough data for both measurand areas are in \ref{appB}, table~\ref{tab:B:H} and~\ref{tab:B:Sq}, after the post-processing of the micrographs and value extraction \cite{euspen2015}.

\begin{figure}[ht]
\centering
\begin{minipage}[c]{.5\textwidth}
    \centering
\includegraphics[width=77mm]{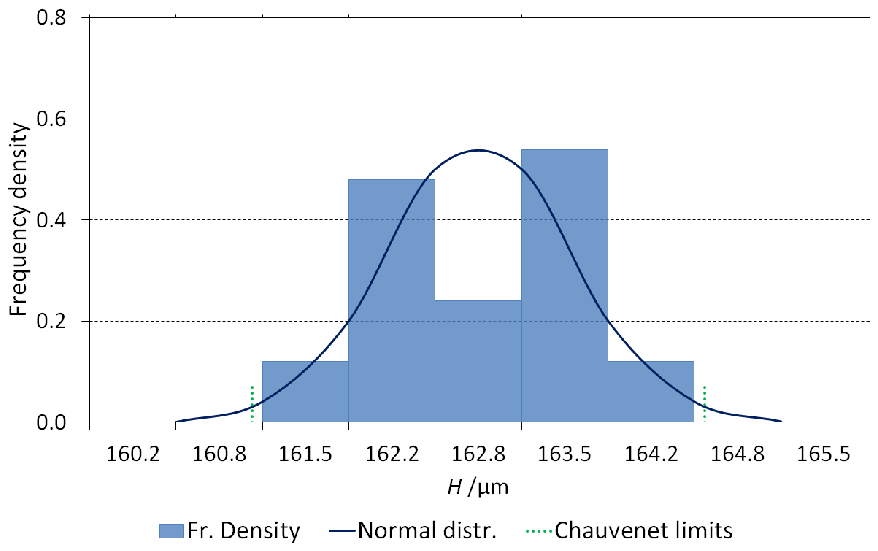}\\
		\vspace{5mm} (a)
\includegraphics[width=77mm]{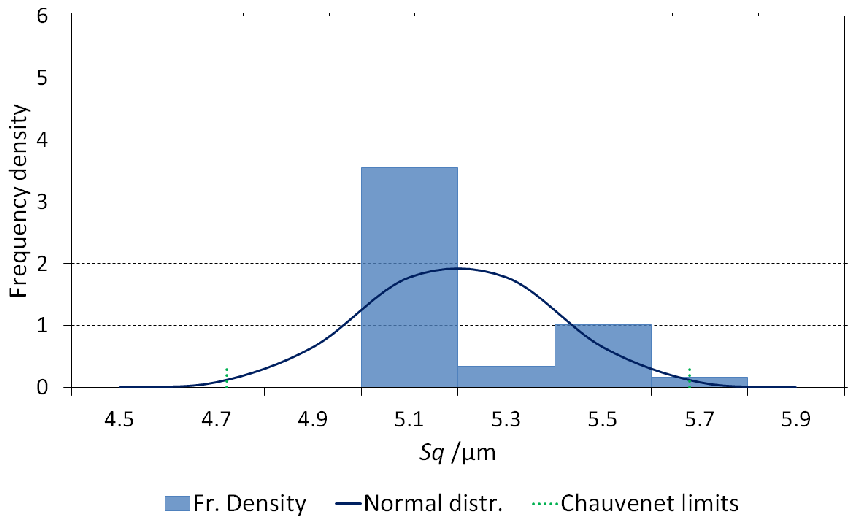}\\
		\vspace{5mm} (c)
  \end{minipage}%
  \begin{minipage}[c]{.5\textwidth}
    \centering
\includegraphics[width=77mm]{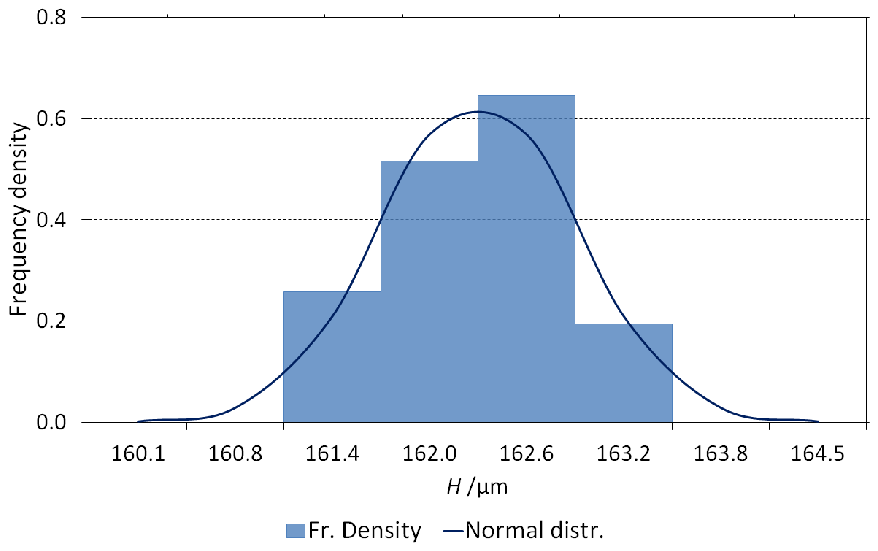}\\
		\vspace{5mm} (b)
\includegraphics[width=77mm]{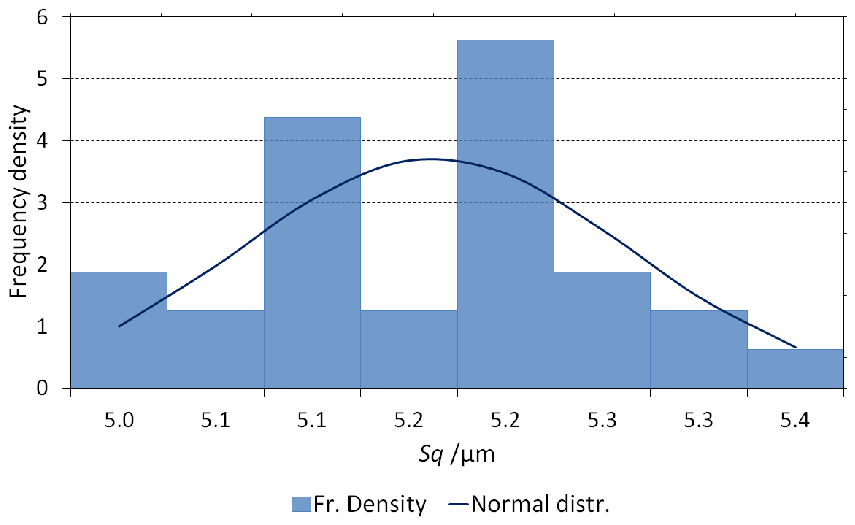}\\
		\vspace{5mm} (d)
  \end{minipage}
	\caption{\label{fig:hist2}Histograms of the experimental distribution, and comparable curves of the normal distribution for the  step height $H$ of (before (a)---Chauvenet's exclusion limits $x_{L~min}=161.1{\rm~\mu m}$, $x_{L~max}=164.6{\rm~\mu m}$---and after (c) the correction of the systematic error), and for the RMS values $Sq$ (before (b)---Chauvenet's exclusion limits $x_{L~min}=4.7{\rm~\mu m}$, $x_{L~max}=5.7{\rm~\mu m}$---and after (d) the correction of the systematic error).}
\end{figure}

Despite the Chauvenet’s criterion did not evidence any outlier, inspecting the blox-plot representations both measurands show only reduced agreement, being it mostly for measurements from $20\times$ and $50\times$ lens objectives.
Moreover, the histograms in figure~\ref{fig:hist2}-a and~-c demonstrated non-normal multi-modal trend of the experimental distributions. This was confirmed by the general linear model ANOVA test in table~\ref{tab:ANOVA2} (sequential sum of squares), where both the \textit{day of acquisition} and the \textit{magnification} were influence factors. The acquisitions, in fact, were completed in six days, three per each specimen. The results of the ANOVA test indicated that both influence factors were significant for both measurands.

\begin{figure}[ht]
\centering
\begin{minipage}[c]{.5\textwidth}
    \centering
    \includegraphics[width=75mm]{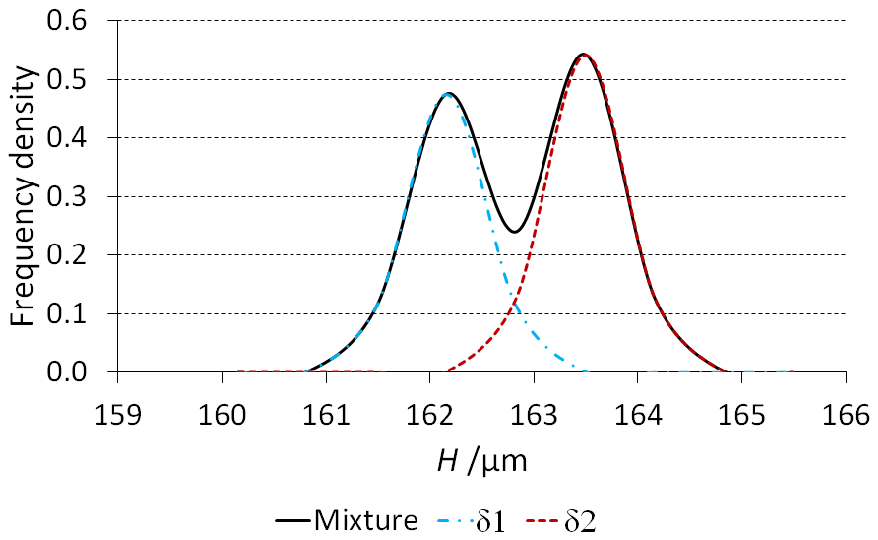}\\
		\vspace{5mm} (a)
    \includegraphics[width=75mm]{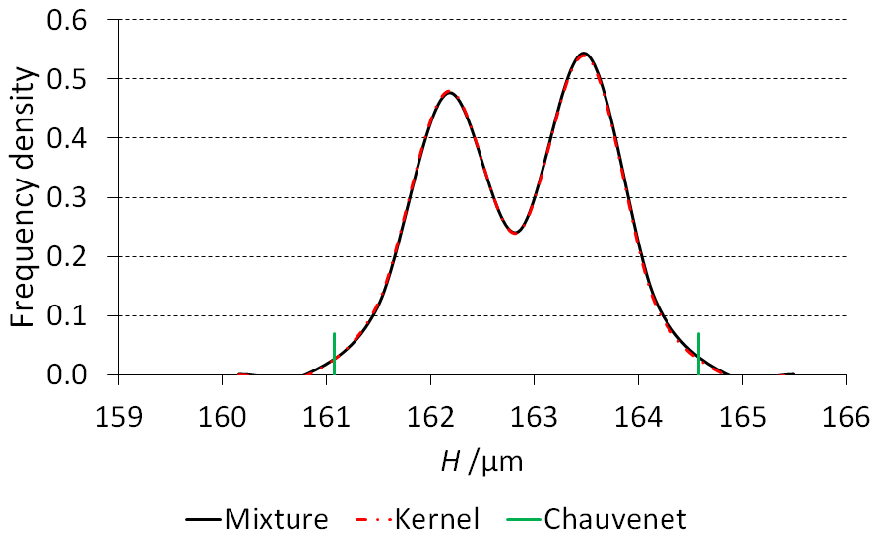}\\
		\vspace{5mm} (c)
  \end{minipage}%
  \begin{minipage}[c]{.5\textwidth}
    \centering
    \includegraphics[width=75mm]{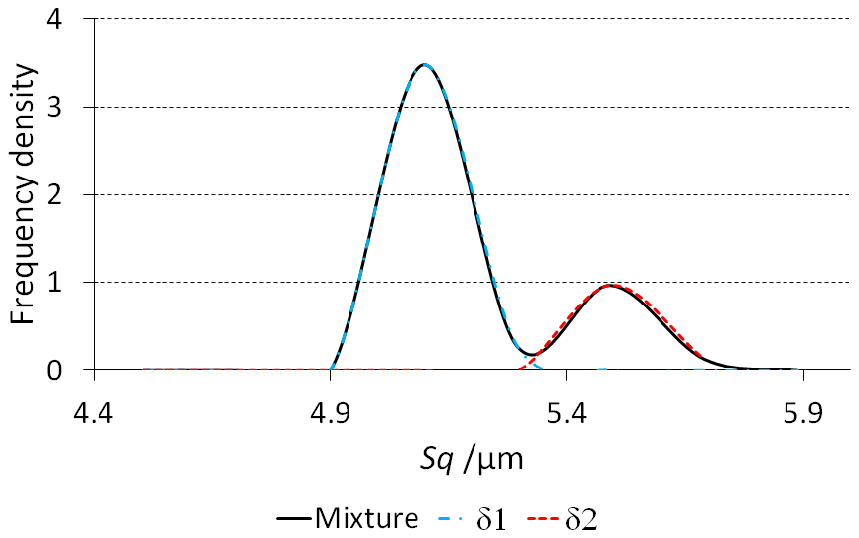}\\
		\vspace{5mm} (b)
    \includegraphics[width=75mm]{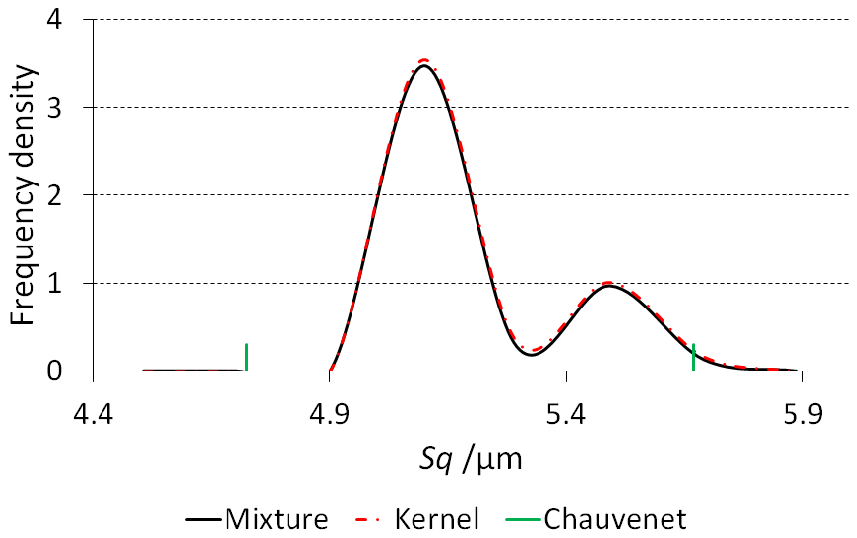}\\
		\vspace{5mm} (d)
  \end{minipage}
  \caption{\label{fig:mixture2}Comparison between mixture and estimation of the probability density of the experimental data (kernel). (a) Mixture of step height $H$. (b) Mixture of RMS values $Sq$. (c) Comparison between kernel and mixture of step height $H$. (d) Comparison between kernel and mixture of RMS values $Sq$.}
\end{figure}

A KDE successively confirmed the presence of two dominant influence factors, and allowed to identify the associated individual kernels by comparison with sub-group averages. They are shown in figure~\ref{fig:mixture2}, while the corresponding parameters are in the tables~\ref{tab:mixture2:H} and~\ref{tab:mixture2:Sq}.

\begin{figure}[ht]
\centering
\begin{minipage}[c]{.5\textwidth}
    \centering
    \includegraphics[width=75mm]{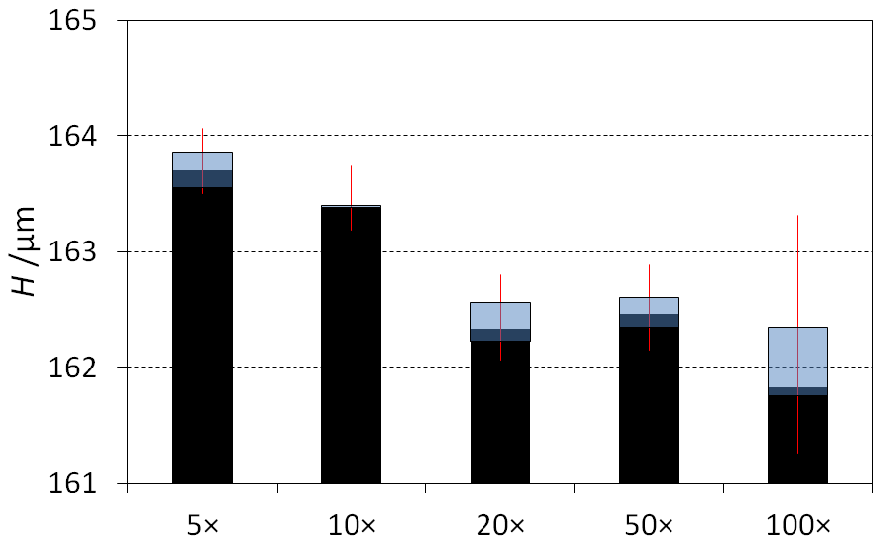}\\
		\vspace{5mm} (a)
  \end{minipage}%
  \begin{minipage}[c]{.5\textwidth}
    \centering
    \includegraphics[width=75mm]{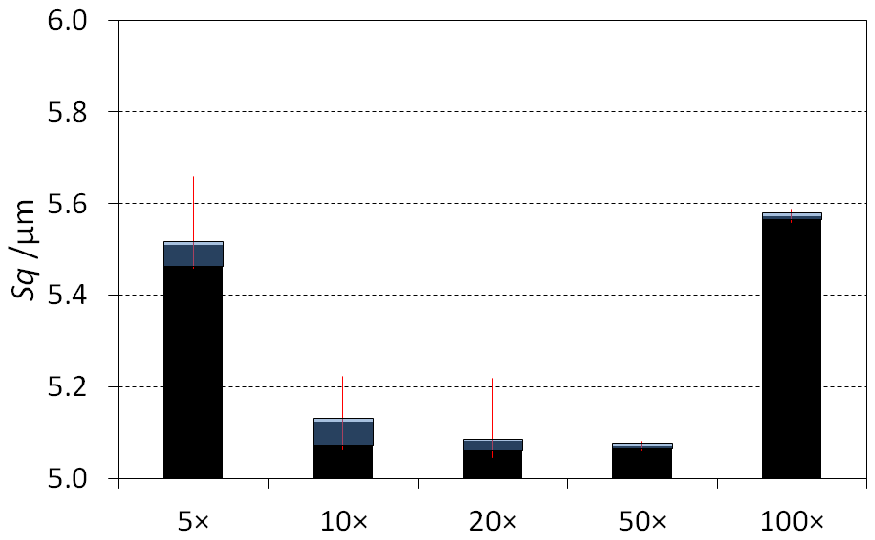}\\
		\vspace{5mm} (b)
  \end{minipage}
  \caption{\label{fig:boxplot2}Box-plots of measurands with interquartile range (box), maximum (upper whisker), minimum (lower whisker) and median (column in the box). (a) Step height $H$ of measurand area A13 (Chauvenet's exclusion limits $x_{L~min}=161.1{\rm~\mu m}$, $x_{L~max}=164.6{\rm~\mu m}$). (a) RMS values $Sq$ of measurand area \textsl{A21} (Chauvenet's exclusion limits $x_{L~min}=4.7{\rm~\mu m}$, $x_{L~max}=5.7{\rm~\mu m}$).}
\end{figure}   

\begin{table}[ht]
\caption{\label{tab:ANOVA2}General linear model ANOVA of measurand area \textsl{A13} and measurand area \textsl{A21} (sequential sum of squares---confidence level 95~\%).}
\footnotesize
\begin{indented}
\lineup
\item[]\begin{tabular}{@{}lll}
\br
							& $H$                        & $Sq$\\
\mr
Day       		& Influence	(p-val $< 0.01$) & Influence (p-val $< 0.01$)\\
Magnification	& Influence (p-val $< 0.01$) & Influence (p-val $< 0.01$)\\
$R^{2}$				& 76~\%                      & 91~\%\\
\br
\end{tabular}\\
\end{indented}
\end{table}
\normalsize

\begin{table}[ht]
\caption{\label{tab:mixture2:H}Parameters of the individual kernels in the mixture related to measurand area \textsl{A13} (optimised $\chi^2$ statistics is 0.62).}
\footnotesize
\begin{indented}
\lineup
\item[]\begin{tabular}{@{}lllll}
\br
Dominant factor & Individual ker. & Average${\rm~/\mu m}$ & St. Dev.${\rm~/\mu m}$ & Percent. Of incid.~/\%\\
\mr
Day           & $\delta_1$ & $162.16$ & $0.40$ & $47$\\
Magnification & $\delta_2$ & $163.49$ & $0.39$ & $53$\\
\br
\end{tabular}\\
\end{indented}
\end{table}
\normalsize

\begin{table}[ht]
\caption{\label{tab:mixture2:Sq}Parameters of the individual kernels in the mixture related to measurand area \textsl{A21} (optimised $\chi^2$ statistics is 3.11).}
\footnotesize
\begin{indented}
\lineup
\item[]\begin{tabular}{@{}lllll}
\br
Dominant factor & Individual ker. & Average${\rm~/\mu m}$ & St. Dev.${\rm~/\mu m}$ & Percent. Of incid.~/\%\\
\mr
Day           & $\delta_1$ & $5.15$ & $0.06(0)$ & $76.5$\\
Magnification & $\delta_2$ & $5.55$ & $0.06(3)$ & $23.5$\\
\br
\end{tabular}\\
\end{indented}
\end{table}
\normalsize

Unlike the previous case of section~\ref{reference}, in this instance the reference was not used in the regression, and the correction of the systematic behavior was performed with respect to the time sequence $t_s$ of acquisition of the repeated measurements. Afterwords, the traceability was achieved as average distance of the experimental distribution from the average trend of the corresponding contact measurements $x_{CI}$. Thus, the metrological model equation giving the values corrected for the systematics was

\begin{equation}
y_{OPT} = x_{CI} + x_{OPT} - x_{regr} \pm \epsilon_{Rep} \pm \epsilon_{res}
\label{met_eq2}
\end{equation}

where $x_{OPT}$ are the optical measurements, and $x_{regr}$ is the mathematical model equation found consistent for fitting the experimental distributions. A first order polynomial was found for the measurand area \textsl{A13}: $x_{regr~H} = a + b \cdot t_s$; and a second order one for the measurand area \textsl{A21}: $x_{regr~Sq} = a + b \cdot t_s + c \cdot t^{2}_{s}$. The related parameters are in table~\ref{tab:regression2}.

For the sake of clarity, it should be noted that (\ref{met_eq}) is different from (\ref{met_eq2}). The first one, if inverted, gives an estimate of a reference measurement, and the related uncertainty is consequence of the regression model. Equation~(\ref{met_eq2}), instead, achieves the correction of the systematic behavior as function of the time sequence $t_s$, and the accuracy as \textsl{distance} from the reference (cf. also with (\ref{met_eq3}) in section~\ref{equo}, and (\ref{met_eq4:1}) in section~\ref{charact}).

Similarly to section~\ref{reference}, the standard uncertainty was calculated propagating the uncertainty contributors to (\ref{met_eq2}) by the usual method of combination of variances for uncorrelated quantities, where the contributors were as follows:

\begin{itemize}
\item	The accuracy of CI, using the values $H_{CI} = (162.30 \pm 0.11)\rm~\mu m$ and $Sq_{CI} = (5.19 \pm 0.11)\rm~\mu m$ from the previous section~\ref{reference}.
\item	The standard deviation of the coefficient of the model equation assessed in the best fit regression (see table~\ref{tab:regression2}).
\item	The standard deviation of the residuals (see table~\ref{tab:regression2}).
\item	The post-processing software numerical precision (estimated to be 1~nm---conservative choice).
\end{itemize}

Finally, the expanded uncertainty was evaluated as the confidence interval corresponding to the conventional confidence level of 95~\%, and by a coverage factor calculated using the $t$-distribution with degrees of freedom given by the Welch-Satterthwaite formula \cite{GUM}.

\begin{figure}[ht]
\centering
\begin{minipage}[c]{.5\textwidth}
    \centering
\includegraphics[width=77mm]{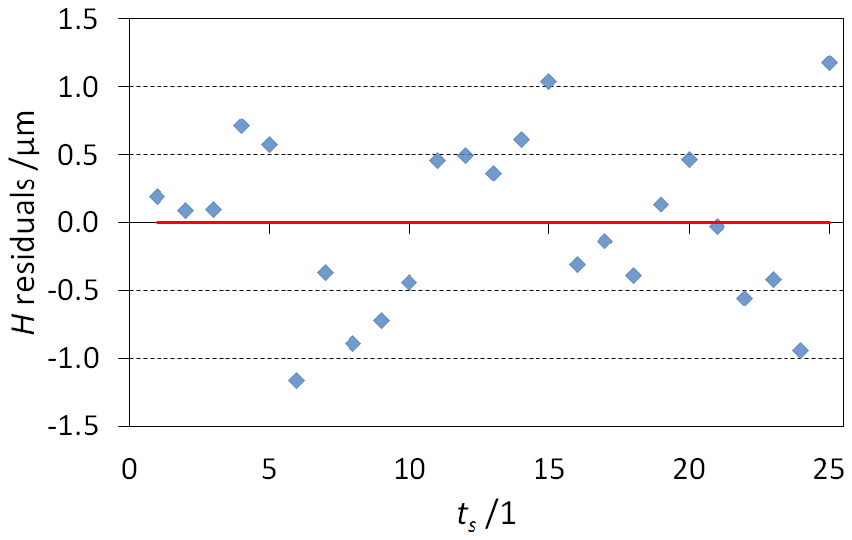}\\
		\vspace{5mm} (a)
  \end{minipage}%
  \begin{minipage}[c]{.5\textwidth}
    \centering
\includegraphics[width=77mm]{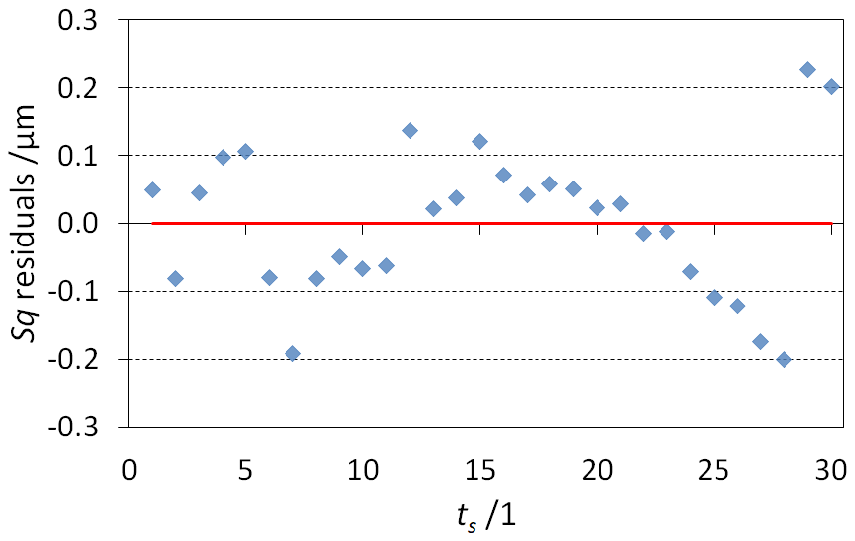}\\
		\vspace{5mm} (b)
  \end{minipage}
	\caption{\label{fig:regression2}Residuals related to measurand area \textsl{A13} (a), and to measurand area \textsl{A21} (b), both represented as function of the acquisition sequence. The straight line indicate the tendency (first-order polynomial).}
\end{figure}

The results are summarized in the tables~\ref{tab:results1} and~\ref{tab:results2} for the sub-groups of influence factors. Moreover, the residuals of the correction are in figure~\ref{fig:regression2}. The residuals' trend indicated an effective correction (no tendency in the residuals' distributions). Nonetheless, inspecting the histograms of the same distributions after correcting for the systematics (see figure~\ref{fig:hist2}-b and~-d) a partial compensation of the systematic behavior was observed in both cases. It was judged not acceptable for the measurand area \textsl{A21}.

\begin{table}[ht]
\caption{\label{tab:regression2}Parameters of the least square regressions. Intercept $a$, slope $b$, quandratic term $c$, standard deviation of the intercept $\sigma_a$, standard deviation of the slope $\sigma_b$, standard deviation of the quandratic term $\sigma_c$, reproducibility $\epsilon_{Rep}$ (standard deviation of the residuals), degrees of freedom and coefficient of determination $R^2$.}
\footnotesize
\lineup
\begin{tabular}{@{}llllllllll}
\br
          &  $a{\rm~/\mu m}$ & $b{\rm~/\mu m}$ & $c{\rm~/\mu m}$ & $\sigma_a{\rm~/\mu m}$ & $\sigma_b{\rm~/\mu m}$ & $\sigma_c{\rm~/\mu m}$ & {$\epsilon_{Rep}{\rm~/\mu m}$} & {DoF} & {$R^2~/$\%}\\
\mr
$x_{regr~H}$  & $162.83$ & $\-0.058$ & --      & $0.13$ & $0.02$  & --       & $0.63$ & $23$ & $32$\\
$x_{regr~Sq}$ & $5.03$   & $0.01$   & $0.002$ & $0.03$ & $0.002$ & $0.0003$ & $0.11$ & $27$ & $70$\\
\br
\end{tabular}\\
\end{table}
\normalsize

\begin{table}[ht]
\caption{\label{tab:results1}Step height results for the measurand \textsl{A13} after the correction. The values are for the sub-groups of influence factors, and related expanded uncertainties.}
\footnotesize
\lineup
\begin{tabular}{@{}lllllllll}
\br
&$5\times$&$10\times$&$20\times$&$50\times$&$100\times$&Day~1&Day~2&Day~3\\
\mr
${H{\rm~/\mu m}}$&$162.63$&$162.89$&$161.58$&$162.25$&$162.14$&$162.38$&$162.32$&$162.11$\\
${U{\rm~/\mu m}}$&$1.41$&$1.38$&$1.37$&$1.38$&$1.41$&$1.41$&$1.38$&$1.41$\\
\br
\end{tabular}\\
\end{table}
\normalsize

\begin{table}[ht]
\caption{\label{tab:results2}RMS results for the measurand \textsl{A21} after the correction. The values are for the sub-groups of influence factors, and related expanded uncertainties.}
\footnotesize
\begin{indented}
\lineup
\item[]\begin{tabular}{@{}lllllllll}
\br
&$5\times$&$10\times$&$20\times$&$50\times$&$100\times$&Day~1&Day~2&Day~3\\
\mr
${Sq{\rm~/\mu m}}$&$5.23$&$5.10$&$5.21$&$5.07$&$5.40$&$5.18$&$5.22$&$5.16$\\
${U{\rm~/\mu m}}$ &$0.45$&$0.44$&$0.44$&$0.45$&$0.46$&$0.44$&$0.44$&$0.45$\\
\br
\end{tabular}\\
\end{indented}
\end{table}
\normalsize

\subsection{Equalization of the micrographs' quantization level}\label{equo}
As a consequence of the results in measurand area \textsl{A21} of section~\ref{t-seq}, the corresponding micrographs were all reviewed. The original measurements were matrices of several fields of view, sized to have approximately similar areas, but still heterogeneous. Taking care of the alignment, they were all re-sized to an area of $9320\times7200$ pixels, with pixel size of about $0.193{\rm~\mu m}\times0.181{\rm~\mu m}$. Thus, the analysis was repeated on the equalized micrographs in the same fashion as in section~\ref{t-seq}. The corresponding raw data are in \ref{appB}, table~\ref{tab:B:SqBand}, after post-processing and value extraction.

The Chauvenet's criterion indicated four outliers, two in the sub-group of $10\times$ magnification and other two in the sub-group $100\times$ magnification. The outliers were eliminated and replaced by the median of all the other values.

\begin{figure}[ht]
\centering
\includegraphics[width=77mm]{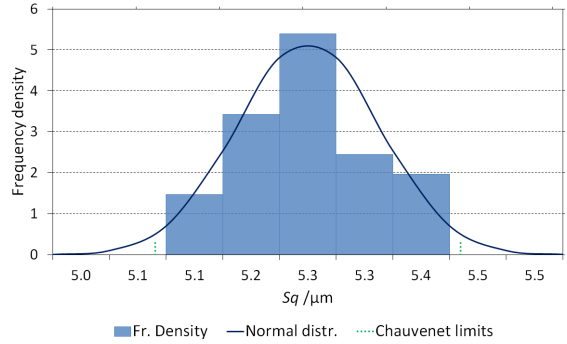}\\
\vspace{2mm} 
\caption{\label{fig:hist3}Histogram of the experimental distribution, and comparable curve of the normal distribution for the measurand area \textsl{A21}, after the equalization of the quantization level in the corresponding micrographs and outliers removal (Chauvenet's exclusion limits $x_{L~min}=5.1{\rm~\mu m}$, $x_{L~max}=5.4{\rm~\mu m}$).}
\end{figure}

The histogram of the experimental distribution in figure~\ref{fig:hist3} indicated a less uneven shape. Although, the results of the ANOVA test confirmed the acquisition day and, surprisingly, the magnification as significant influence factors (see table~\ref{tab:ANOVA3}).

\begin{figure}[ht]
\centering
\includegraphics[width=77mm]{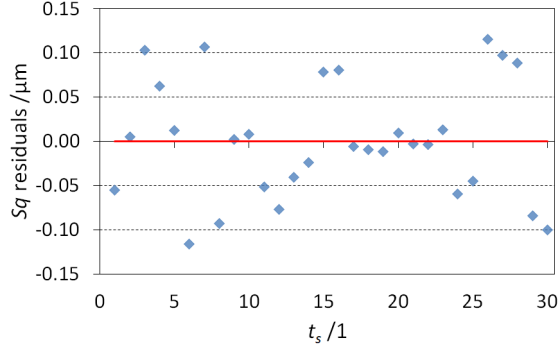}\\
\vspace{2mm}
\caption{\label{fig:regression2band}Residuals related to measurand area \textsl{A21} with equalized quantization level. The straight line indicate the tendency (first-order polynomial).}
\end{figure}

The parameters of the regression for the correction are in tables~\ref{tab:regression3}, while the results for the sub-groups of influence factors are in table~\ref{tab:results3}.
Comparing figure~\ref{fig:regression2band} with figure~\ref{fig:regression2}-b, it is worth to note that the variability of the residuals (and, thus, the reproducibility) was reduced to about half range in consequence of the equalization.

\begin{table}[ht]
\caption{\label{tab:ANOVA3}General linear model ANOVA of measurand area \textsl{A21} (sequential sum of squares---confidence level 95~\%).}
\footnotesize
\begin{indented}
\lineup
\item[]\begin{tabular}{@{}ll}
\br
							& $Sq$\\
\mr
Day       		& Influence (p-val $\approx 0.01$)\\
Magnification	& Influence (p-val $< 0.01$)\\
$R^{2}$				& 69~\%\\
\br
\end{tabular}\\
\end{indented}
\end{table}
\normalsize

\begin{table}[ht]
\caption{\label{tab:regression3}Parameters of the least square regression. Intercept $a$, slope $b$, quandratic term $c$, standard deviation of the intercept $\sigma_a$, standard deviation of the slope $\sigma_b$, standard deviation of the quandratic term $\sigma_c$, reproducibility $\epsilon_{Rep}$ (standard deviation of the residuals), degrees of freedom and coefficient of determination $R^2$.}
\footnotesize
\lineup
\begin{tabular}{@{}llllllllll}
\br
          &  $a{\rm~/\mu m}$ & $b{\rm~/\mu m}$ & $c{\rm~/\mu m}$ & $\sigma_a{\rm~/\mu m}$ & $\sigma_b{\rm~/\mu m}$ & $\sigma_c{\rm~/\mu m}$ & {$\epsilon_{Rep}{\rm~/\mu m}$} & DoF & {$R^2~/$\%}\\
\mr
$x_{regr~Sq}$ & $5.21$   & $\-0.0001$   & $0.001$ & $0.02$ & $0.0001$ & $0.0002$ & $0.07$ & $27$ & $24$\\
\br
\end{tabular}\\
\end{table}
\normalsize

\begin{table}[ht]
\caption{\label{tab:results3}RMS results for the measurand \textsl{A21} after the correction of the equalized micrographs. The values are for the sub-groups of influence factors, and related expanded uncertainties. The decimals in parentheses explicate the approximation on the last digit.}
\footnotesize
\begin{indented}
\lineup
\item[]\begin{tabular}{@{}lllllllll}
\br
&$5\times$&$10\times$&$20\times$&$50\times$&$100\times$&Day~1&Day~2&Day~3\\
\mr
${Sq{\rm~/\mu m}}$&$5.22$&$5.17$&$5.18$&$5.31$&$5.10$&$5.18$&$5.20$&$5.20$\\
${U{\rm~/\mu m}}$ &$0.40(9)$&$0.40(7)$&$0.40(9)$&$0.41(1)$&$0.41(4)$&$0.40(6)$&$0.40(9)$&$0.41(3)$\\
\br
\end{tabular}\\
\end{indented}
\end{table}
\normalsize

\subsection{Traceability by material measures}\label{trac}
The use of a reference instrument for establishing the traceability may lead to the undesired possible presence of additional influence factors. 
A prevailing solution for the achievement of the traceability in manufacturing is through measurement standards that are \textit{material measures} \cite{VIM}. Measuring calibrated material measures with measurands' dimensions very close to the quantities under investigation allows for the estimation of the instrument's performance, thus, establishing the traceability. This was already noticed in the Introduction, concerning the metrological characteristics.

Differences from the calibrated values can also be compensated in the model equation by the \textit{substitution method}. For instance, the substitution method for uncertainty evaluation, based on material measures and concerning coordinate measuring machines, is described in the GPS standard ISO~15530-3 \cite{ISO15530-3}.

The use of calibrated material measures implies two sessions of measurements with the same surface topography measuring instrument: measurements of the material measure $x_{ref,gauge}$ and measurements of the components under characterization $x_{OPT}$. Comparing $x_{ref,gauge}$ with the calibrated values in the calibration certificate $x_{ref,cal} \pm U_{ref,cal}$, the traceability for the characterization measurements can be achieved using a calibration factor given by the ratio $F_{cal}=x_{ref,cal}/x_{ref,gauge}$. Supposing to correct the systematics for both sets of measurements (same number of repeated measurements performed) with a first-order polynomial, namely $x_{regr,ref,gauge}=q_g \cdot t_s \pm \epsilon_{Rep,gauge}$ and $x_{regr,OPT}=p_{OPT} \cdot t_s \pm \epsilon_{Rep,OPT}$ the metrological model equation is 

\begin{eqnarray}
y_{OPT} = F_{cal} \cdot x_{regr,OPT} = \frac{x_{ref,cal}}{x_{regr,ref,gauge}} \cdot x_{regr,OPT}=\nonumber\\
= x_{ref,cal} \cdot \frac{p_{OPT} \cdot t_s \pm \epsilon_{Rep,OPT}}{q_g \cdot t_s \pm \epsilon_{Rep,gauge}}.\label{met_eq3} 
\end{eqnarray}

The uncertainty contributors to be referred to (\ref{met_eq3}) are

\begin{itemize}
\item	The uncertainty stated in the calibration certificate of the material measure $U_{ref,cal}$.
\item	The standard deviations of the coefficients in the best fit regressions ($x_{regr,ref,gauge}$, $x_{regr,OPT}$).
\item	The standard deviations of the residuals of the best fit regressions ($\epsilon_{Rep,OPT}$, $\epsilon_{Rep,OPT}$).
\item Uncertainty contributors of other possible influence factors not in the model equation (random variables with null average).
\end{itemize}

\section{Correction of the systematic behavior in surface characterization}\label{charact}
The results of the previous case showed a significant reduction of the measurement uncertainty as a consequence of the correction of the systematics, although the elimination of the influence factors was not complete. Therefore, aim of the last instance was to test efficacy and generality of the measurement uncertainty possible reduction in consequence of such correction.

The correction of the systematic behavior was implemented on the root mean square $Sq$ of the height values of the topographic measurements of four different polished surfaces \cite{ISO25178-2}. The approach was validated comparing the associated measurement uncertainties with the one of the same measurements and same uncertainty contributors assessed without correction.

The four surfaces were polished as follows \cite{4M-1c}:
\begin{itemize}
\item	T1: grade \#15 diamond buff (nominal \textit{Ra} interval 51--76~nm).
\item	T2: 320 grit paper (nominal \textit{Ra} interval 229--254~nm).
\item	T3: 400 stone (nominal \textit{Ra} interval 254--305~nm).
\item	T4: glass bead \#11 dry blast (nominal \textit{Ra} interval 635--711~nm).
\end{itemize}

\begin{figure}[ht]
\centering
\begin{minipage}[c]{.5\textwidth}
    \centering
    \includegraphics[width=62mm]{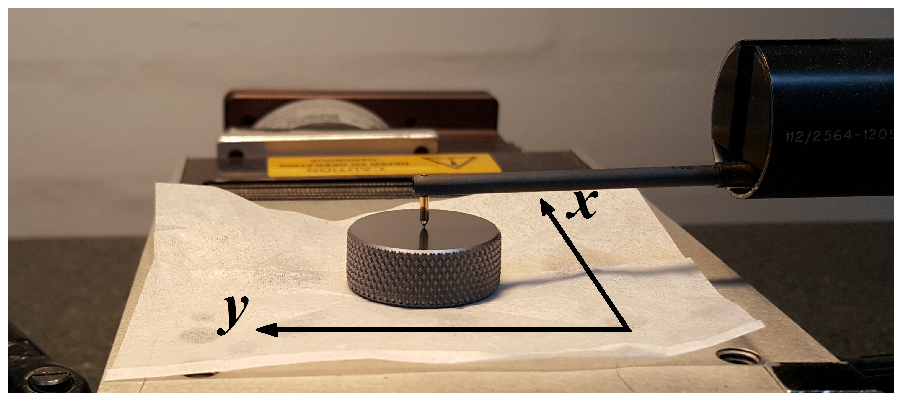}\\
		\vspace{4mm} (a) \vspace{1mm}\\
    ~~~\includegraphics[width=67mm]{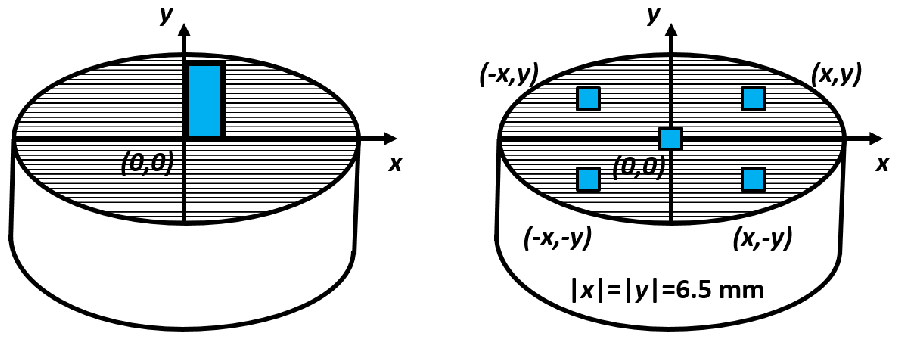}\\
		\vspace{4mm} (b)~~~~~~~~~~~~~~~~~~~~~~~(c)
  \end{minipage}%
  \begin{minipage}[c]{.5\textwidth}
    \centering
    \includegraphics[width=85mm]{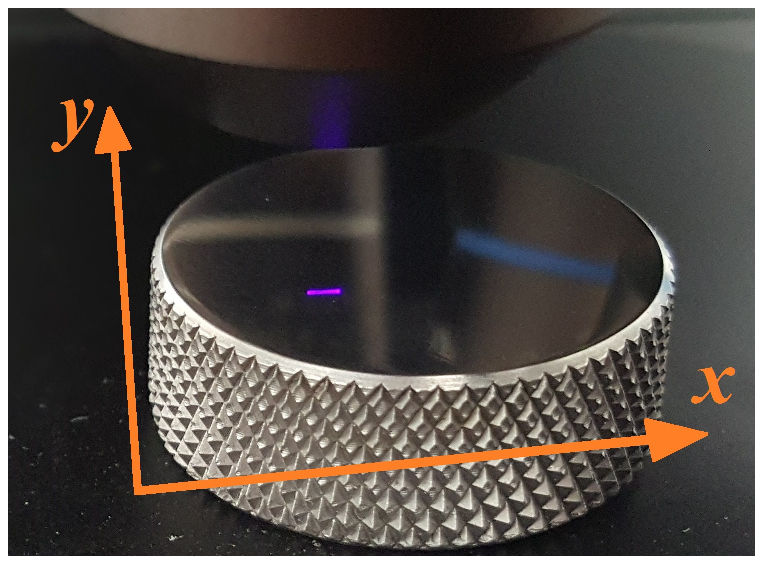} \vspace{1mm}\\
		\vspace{3mm} (d)
  \end{minipage}
  \caption{\label{fig:meaSys}Instruments' set-up and on-specimen local reference system. (a) Specimen on CI and local reference system. (b) Local reference system for CI. (c) Local reference system for LSC. (d) Specimen on LSC and local reference system.}
\end{figure}

At first, they were measured by a reference CI according to the on-sample reference system shown in figure~\ref{fig:meaSys}-a~and~-b, where the dominant texture was oriented orthogonally to the scanning direction ($y$-axis). Five acquisitions of a single evaluation area of $4~{\rm mm}\times10~{\rm mm}$ were performed on each surface, sampled at 8200 pixels along the $x$-axis, and 21 profiles along the $y$-axis (nominal tip radius $2~{\rm\mu m}$). Considering the ranges of interest, the expanded uncertainties stated in the calibration certificate of the CI were 10~nm for $Ra$ values below 229~nm, and 24~nm in the $Ra$ interval 229--604~nm \cite{4M-1c}--\cite{Talysurf} (rough data in \ref{appC}, table~\ref{tab:C:CI}).

The surfaces were successively characterized by a LSC with $100\times$ magnification (quantization level of $4096\times4096$ pixels, with pixel width of about 32~nm), and according to the local reference system and sampling scheme respectively in figure~\ref{fig:meaSys}-c~and~-d (laser scanning movement along the $y$-axis and orthogonal to the dominant texture). Being the surfaces associated with four cylindrical steel tools ($20~{\rm mm}\times10~{\rm mm}$), the origin of the local reference system was set in the center of the cylinders as shown in figure~\ref{fig:meaSys}-c. Thus, ten repeated measurements were performed, in the central position (0,0), and in the peripheral positions (x,y), (x,-y), (-x,y), (-x,-y), where $\left|x\right|=\left|y\right|=6.5$~mm (rough data in \ref{appC}, table~\ref{tab:C:LSC}).

The characterization of a surface normally implies the quantification of functional behaviors in terms of surface texture, which are to be proven on areas that are usually quite large with respect to the field of view of the common surface topography measuring instruments. Therefore, a common choice is the acquisition of larger surface portions as matrices of fields of view (\textit{stitching}). Such approach is useful when large features are to be characterized dimensionally. In this case, instead, the choice was to sample the surfaces at homogeneous spacing because of their expected uniformity. Thus, the investigation intended to average the assessment of the surface texture, so that possible variations due to dissimilarities or irregularities of the surface polishing among different areas could be included in the measurement uncertainty after compensating for the influence of the instrument. Hence, a larger area was chosen for the CI, not covering any of the areas chosen for the LSC to achieve a denser overall surface sampling.

\subsection{Results}
\begin{figure}[ht]
\centering
\begin{minipage}[c]{.5\textwidth}
    \centering
    \includegraphics[width=75mm]{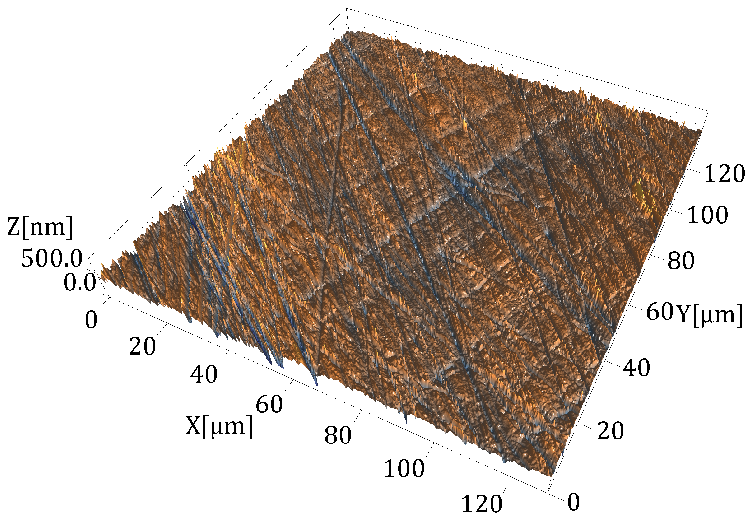}\\
		\vspace{5mm} (a)
    \includegraphics[width=75mm]{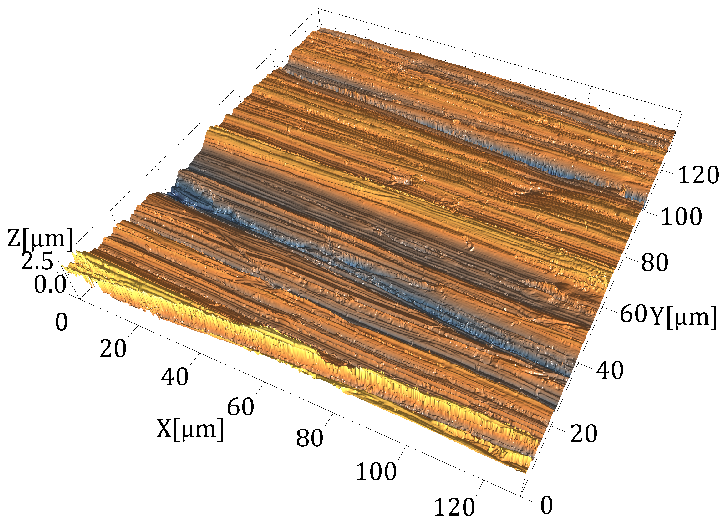}\\
		\vspace{5mm} (c)
  \end{minipage}%
  \begin{minipage}[c]{.5\textwidth}
    \centering
    \includegraphics[width=75mm]{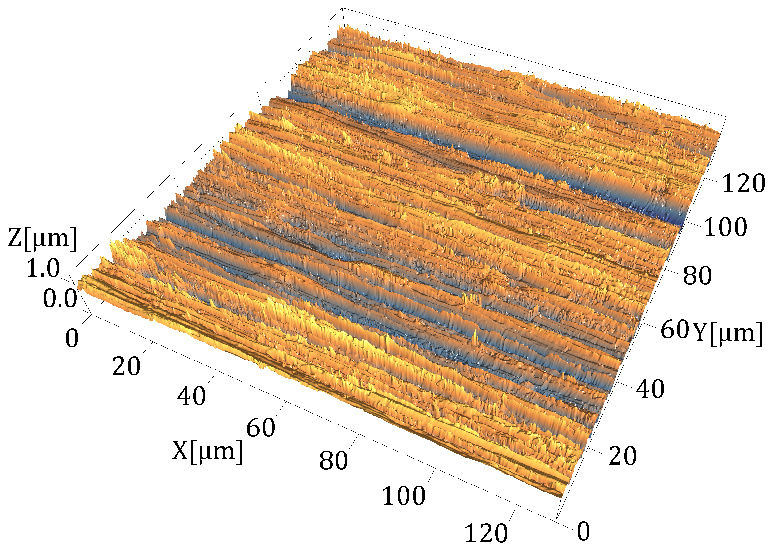}\\
		\vspace{5mm} (b)
    \includegraphics[width=75mm]{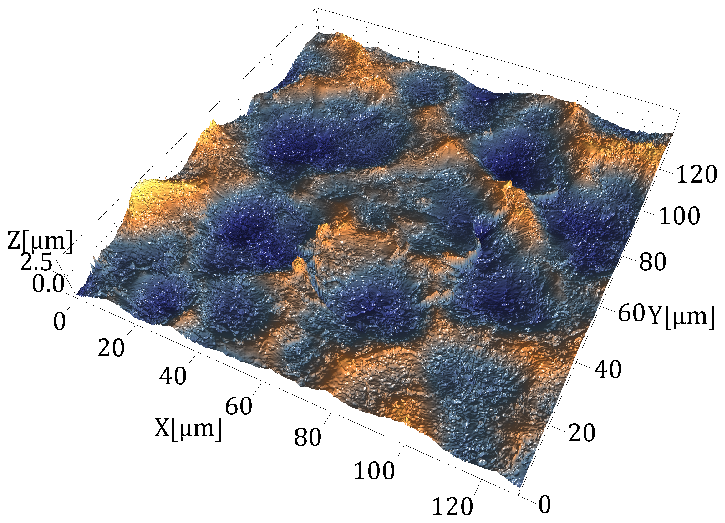}\\
		\vspace{5mm} (d)
  \end{minipage}
  \caption{\label{fig:DME}Examples of acquired micrographs (3D views). (a) T1: diamond buff polishing (grade \#15). (b) T2: 320 grit paper polishing. (c) T3: 400 stone polishing. (d) T4: dry blast polishing (glass bead \#11).}
\end{figure}

The data set of root mean square values \textit{Sq} was examined for outliers by the Chauvenet’s criterion. Four outliers related to T3 were removed, one in each peripheral area (see \ref{appC}---values in parenthesis in table~\ref{tab:C:LSC}).
Examples of each polished surface are in figure~\ref{fig:DME} (3D views of micrographs).

The correction of tha systematics was carried out by a least square fit of the optical LSC measurements as function of the reference CI measurements, according to what already shown in section~\ref{reference}. The optical measurements corrected for accuracy and systematics are given by

\begin{eqnarray}
\fl y_{OPT} = x_{OPT} + x_{OPT~ave} - x_{regr} \pm \epsilon_{Rep} =\nonumber\\
= x_{OPT} + x_{OPT~ave} - q \cdot x_{CI} \pm \epsilon_{Rep}.\label{met_eq4:1} 
\end{eqnarray}

where $x_{OPT}$ are the optical measurements with average $x_{OPT~ave}$, $x_{CI}$ the reference measurements, and $x_{regr}=q \cdot x_{CI}$ is the mathematical model equation found consistent for best fitting the experimental data (straight line with null constant term). The parameters of the least square fit are in \ref{appC}, table~\ref{tab:regression4}.

The measurement uncertainty evaluated according to (\ref{met_eq4:1}) had the following contributors (the complete uncertainty budget is in table~\ref{tab:C:ucorr})
\begin{itemize}
\item	$u_{repr,fit}$: reproducibility values from the regression (standard deviation of the residuals).
\item	$u_{slope}$: standard deviation of the coefficient of the best fit regression.
\item $u_{repea,opt}$: standard deviation of the optical measurements (repeatability).
\item	$u_{CI}$: accuracy of the CI (see (\ref{met_eq:uCI}) in \ref{appC} for more details).
\end{itemize}

When the correction of the systematic behavior was not performed, the traceability was achieved as the mismatch (\textsl{distance}) between optical and contact measurements, leading to the following equation

\begin{equation}
y_{OPT} = x_{OPT} - (x_{OPT~ave} - x_{CI})
\label{met_eq4:2}
\end{equation}

The measurement uncertainty evaluated according to (\ref{met_eq4:2}) had the following contributors (the complete uncertainty budget is in table~\ref{tab:C:uncorr})
\begin{itemize}
\item	$u_{repr}$: maximum deviation of optical measurements (uniformly distributed).
\item	$u_{repea,opt}$: standard deviations of optical measurements (repeatability).
\item $u_{repea,CI}$: standard deviations of reference measurements (CI) (reference repeatability).
\item	$u_{CI}$: accuracy of the CI (see (\ref{met_eq:uCI}) in \ref{appC} for more details).
\end{itemize}

The contributors $u_{repea,opt}$ and $u_{repr}$ account respectively for the variability of the optical measurements $x_{OPT}$ (measurement process), and for the average of optical measurements $x_{OPT~ave}$ (traceability correction), both in (\ref{met_eq4:2}).

\begin{figure}[ht]
\centering
\includegraphics[width=130mm]{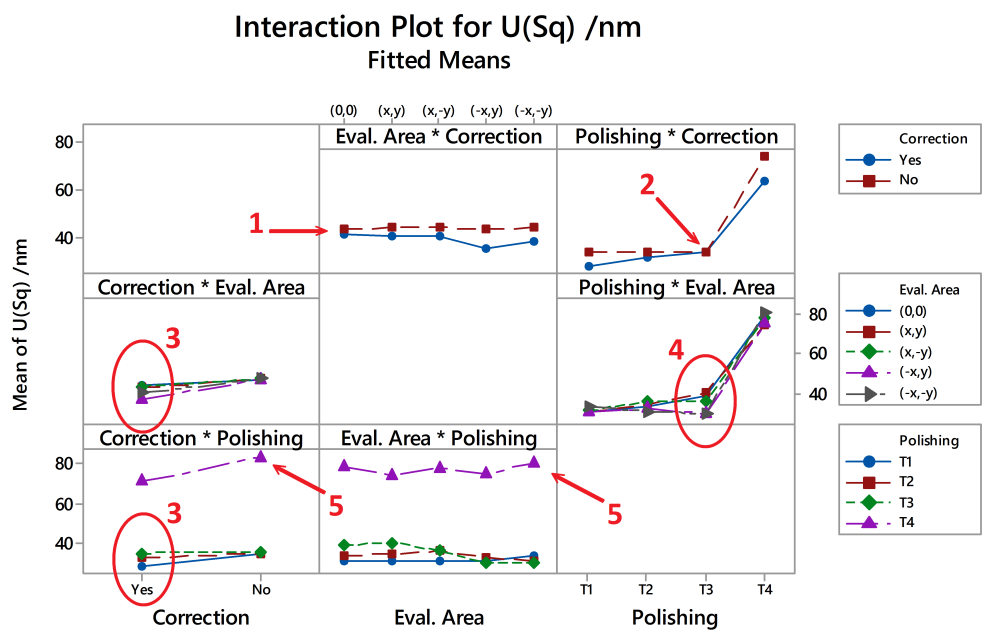}
\vspace{5mm}
\caption{\label{fig:DOE:int}Second-order interaction plot comparing uncertainties related to different measurand areas, with and without correction of the systematic behavior (numbers are referred in the text).}
\end{figure}

Due to the complexity of the investigation and large amount of data, this third instance was summarized by the second-order interaction plot of a general linear model design of experiment (DOE---see figure~\ref{fig:DOE:int}). The DOE was implemented on the entire data set and, also, on partial results corresponding to the following influence factors:
\begin{itemize}
\item	Measurements corrected for the systematic behavior: "Yes", "No".
\item	Evaluation in the central and peripheral areas: "(0,0)", "(x,y)", "(x,-y)", "(-x,y)", "(-x,-y)".
\item	Surfaces with different polishing: "T1", "T2", "T3", "T4".
\end{itemize}

Referring to the numbers in figure~\ref{fig:DOE:int}, the results can be reviewed as follows:
\begin{enumerate}
\item	Systematics' correction in different types of evaluation areas. The evaluated uncertainty was lower after the correction of the systematic effects, regardless of the evaluation area.
\item	Systematics' correction in different types of polishing. When the systematic effects were corrected, a lower uncertainty could be obtained for all the specimens except for T3 (stone polishing). This happened because some of the measurements related to T3 were affected by noise. In fact, they were in the only set of measurements in which an exclusion of outliers was necessary.
\item	Efficacy of the systematics' correction considering different evaluation areas (with respect to the specimen's center) or different types of polishing. The correction produced lower uncertainties with different assessed values depending on the evaluation areas and on the surfaces. Such differences were small, and due to the sought unevenness in the different types of polishing.
\item	Variations due to different sampling areas with respect to the specimen's center. The largest spread of uncertainty values was on the evaluation areas of T3 (presence of noise).
\item	Variations due to different types of polishing. The lowest uncertainty was stated for T1 (diamond buff polishing), whilst the highest was for T4 (dry blast polishing). However, the final uncertainty for T4 was influenced by a higher corresponding traceability uncertainty in the calibration certificate.
\end{enumerate}

\section{Discussion}\label{dis}
A gradual understanding of the bias between frequentist statistics evaluation and the normal practice in the field of metrology for manufacturing was the endeavor of the first instance. The correction of the systematic behavior was a mere attempt, having the most remained unknown because of a general lack of complete information in the measurements data sets. The use of deviations, in fact, put together heterogeneous measurements, where the poorness of repetitions did not allow to discriminate among the three main influence factors highlighted by the KDE. The ANOVA test response identified as influence factors the instrument and the magnification used, but certainly not the measurand area. Thus, a third dominant influence factor remained unknown. The \textit{instrument} influence factor was possibly related to the instrument-operator chain---e.g. different light setting used for the same measurand. While the \textit{magnification} influence factor clearly added differences in the micrographs like the number of pixels and their width, but also other dissimilarities analyzed in the second instance such as the acquisition \textit{day}, which was undeniably the third influence factor. In fact, if the systematic behavior must be considered in its development (see section~\ref{t-seq}), it can be expected starting with the measurement session on a specific instrument and ending when that instrument is turned off. Since optical measurements can be long-lasting, the influence factors must be considered consistently with a measurement session, expecting a possible different behavior of the same factors on other measurement sessions (implying an instrument turned off and on). This was a central point. When comparing different instruments, if the measurements are properly planned with enough repeated measurements, a KDE of a mixture of normal distributions can be exploited to identify the deviations introduced by dominant influence factors. The regression model was also not adequate. The tendencies highlighted in figure~\ref{fig:sequence} could not be corrected, again showing the non-suitability of the measurements planning. The step height measurements would require a second order polynomial. Nevertheless, examining the $x_{devSq}$ distribution and in light of the considerations just made, only a correction by piecewise regressions (i.e. per instruments) would be effective. Thus, different behaviors against one single reference instrument were inconsistent, possibly having the reference a systematic behavior with different trend, too.

Such issues were analyzed more in depth in the second instance, where the measurements were enough to define all the factors influencing the experimental distribution, accounted in their natural sequence of acquisition.

The measurement day as influence factor was a confirmed outcome, being the measurements an inconsistent set inasmuch as some of the conditions responsible for the systematic behavior are restored or changed at the beginning of each measurement session. Hence, for effectively applying the method when there are long lasting acquisitions, the systematics should be related to, and corrected for, continuous measurement sessions planed in homogeneous groups.

The result after the equalization was another interesting outcome. It revealed that the quantization level (or bandwidth \cite{leach4}) is only one side of the problem, being the other one connected with the surface mapping by finite resolution. The content of information was incomplete for $5\times$ (\textit{soft resolution}---where a micrograph does not hold enough details resulting in an extremely low quality of the surface mapping) and $100\times$ (too small area under analysis due to a reduced field of view and, thus, not enough content of information while peculiar details dominate). Instead, it was equivalent for $20\times$ and $50\times$, which suggested they represent the best trade-off quantization--information content for mapping the surface topography (at least for the instrument and the non-smooth surfaces involved---cf. figure~\ref{fig:boxplot2}).
Moreover, the equalization granted more efficacy to the systematics' correction, achieving a final uncertainty significantly reduced (cf. table~\ref{tab:results2} with table~\ref{tab:results3}).

The previous observations were conveyed to the third instance, which proved the correction of the systematics effective in reducing the final uncertainty. Moreover, sampling different areas on the surfaces, and acquiring several repeated measurements for each area allowed for inherently compensating the influence of the instrument, and concurrently stating a measurement uncertainty characterizing the surfaces' unevenness. Specifically, multiple repeated measurements and multiple areal acquisitions together were exploited to separate the variability of the manufacturing process and the one of the instrument (an explicit example of use of both multiplicities can be found elsewhere \cite{CIRP-CAT}).

Nonetheless, optical measurements are relatively long lasting. The validity of the approach is strictly related to the number of repeated measurements that is attainable in an acceptable time. Furthermore, the contact measurements were not adequate as reference. In fact, it emerged clearly that the CI uncertainties had a main impact on the final uncertainty budget, leading to a probable overestimation (cf. uncertainty contributors in table~\ref{tab:C:ucorr}). This was also the case of section~\ref{equo}, where the equalization of the quantization level reduced the spread of the residuals to about half of their variability but the final uncertainty did not reduce consistently because the traceability uncertainty contributor was dominant. Despite the caveat already introduced regarding the artifacts' manufacture (see Introduction), a more adequate choice for establishing the traceability is still the use of material measures (at least, at the time of writing). They are not affected by the limitations inherent in the use of a stylus, viz. an additional systematic behavior, the finite tip radius and an expanded uncertainty inadequate for the present-day demand at the nanoscale.

Regardless of the efficiency on the achievement of the traceability, the proven efficacy of the method on the final uncertainty reduction was also a confirmation of the validity of the starting hypothesis in section~\ref{hyp}.
In other words, average (height of step) and quadratic integral average (root mean square value) of measurands (sets of height values---as explained in section~\ref{hyp}) met the hypotheses of TCL. In fact, even though the systematic behavior $f(x_i)$ could not be recognized completely ($\hat{f}(x_i)$ is an estimate in (\ref{eq:syst:1}) and (\ref{eq:syst:2})), it was possible to correct the experimental distribution to a quasi-random one because the correction had zero expectation (\ref{eq:syst:3}), and the unrecognized systematic effects were \textit{randomized} in the experimental distribution
\numparts
\begin{eqnarray}
\label{eq:syst:1}y_i=x_i+f(x_i)-\hat{f}(x_i)\\
\label{eq:syst:2}f(x_i)-\hat{f}(x_i)\neq 0\\
\label{eq:syst:3}E\left\{f(x_i)-\hat{f}(x_i)\right\}=0.
\end{eqnarray}
\endnumparts

Conversely, failure in the validity of the starting hypotheses of TCL leads to a correction with a non-zero expected value of the \textit{randomized} systematic components. Therefore, the non-zero expected value may become a bias increasing the overall evaluated uncertainty \cite{Pavese-book}.\\

The possible use of averages for dimensional or topographic characterization is rigorously connected with the sought functionality of the component under test. Free-form surfaces are very common nowadays, where the metrology tasks urge more and more the adoption of \textit{distributed measurands}. Although, the instinctive belief of applying frequentist statistics pixel-wise for avoiding averages would raise even more complications because of the nature of the micrographs.

\begin{figure}[t]
\centering
\begin{minipage}[c]{.5\textwidth}
    \centering
		\includegraphics[width=77mm]{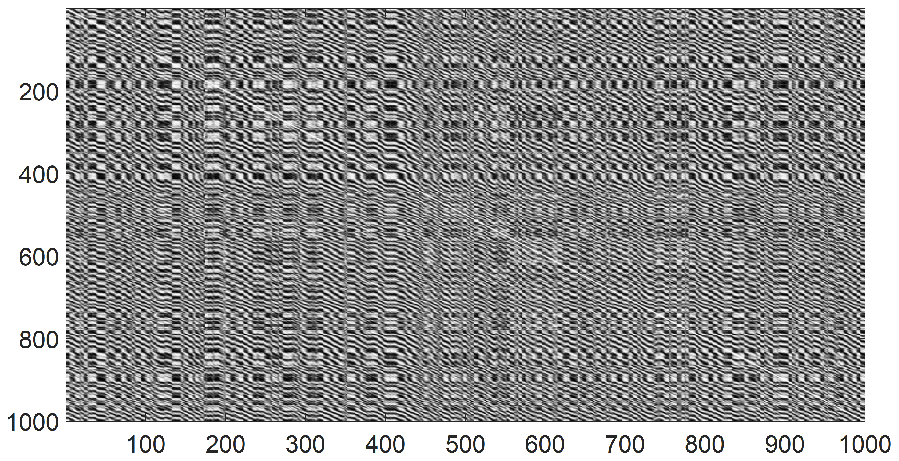}\\
		\vspace{5mm} (a)
  \end{minipage}%
  \begin{minipage}[c]{.5\textwidth}
    \centering
    \includegraphics[width=75mm]{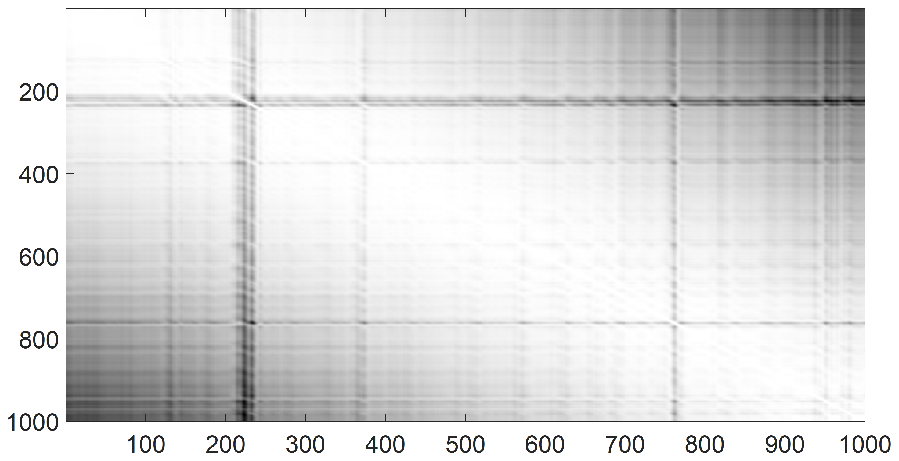}\\
		\vspace{5mm} (b)
  \end{minipage}
	\caption{\label{fig:covmat}Portions of covariance matrices of a lapped surface acquired by CSI using $20\times$ (a) and $50\times$ (b) magnification objectives. The covariance matrices are visualized by a gray-scale palette, where the maximum value is normalized to white and the minimum one to black. Being the micrographs matrices of $1000\times1000$ elements (one million elements each), only the central rows of the respective covariance matrices are visualized in the figure.}
\end{figure}

Figure~\ref{fig:covmat} shows portions of two normalized covariance matrices (Pearson's correlation) as imaged in grey-scale palette, where the achievable maximum +1 and minimum -1 were associated respectively to white and black. The calculation was applied to micrographs of a lapped surface acquired by CSI using $20\times$ (figures~\ref{fig:covmat}-a) and $50\times$ (figures~\ref{fig:covmat}-b) magnification objectives. Being the micrographs matrices of $1000\times1000$ elements (one million elements each), only the central rows of the respective covariance matrices are visualized in the figure.


The correlation appears periodic and complex, with several inversions between positive and negative fluctuating values. The shown covariance also suggests that the correlation between one specific pixel and one of the neighboring pixels fade away after a certain \textit{characteristic} length (mainly depending on the nature of the imaging sensor, and on the degree of excitation of the constituent elements). Whereas now it is clear that the application of the frequentist approach to inference was possible in this work because the correlation was resolved by averaging among height values, and the systematic behavior did not involve interactions among influence factors.
Therefore, the correction of the systematics considering the measurand set of height values demands methods more sophisticated than the GUM approach, which are also required to be robust to noise.
The noise in micrograph acquisitions is, in fact, the cause of different errors affecting the pixels' measure. A typical example is the noise causing a sharp increase in the magnitude of some pixels' measure (in jargon \textsl{spikes}). Spikes are outliers in the distribution of the surface mapped values. Nonetheless, the usual methods for the metrological outliers' elimination would not be capable to resolve such presence, creating indeterminacy. In fact, metrological outliers are modeled statistically as rare events, thus, the removal methods are based on low probability of occurrence. On the contrary, divers types of noise exists that have their own specific distributions (white noise, colored noise, flicker noise, etc.), which bring them closer to a systematic behavior rather than metrological outliers. Moreover, the presence of low-frequency noise (e.g. flicker noise from amplifiers) mixed with high-frequency noise hinders the possibility of elimination e.g. by filtering. A discussion on the effects of the measurement noise is beyond the scope of this work. Nonetheless, the aim is to advise on the application of methods that are consolidated in many fields of metrology, although not verified and validated with optical measurements due to the complexity of the operation of the instruments involved.

\section{Conclusions}\label{conc}
The sequence of the proposed case studies clearly highlighted that precautions are necessary for the application of frequentist statistics to microscopes' acquisitions, planning in advance proper measurement sessions. The first case analyzed \textit{bad habits} in measurements and processing, still quite diffused in the field. The influence factors due to bad practice add to the systematic behavior of the instrument, counteracting the correction. Therefore, bad practice must be avoided to have an effective application of the method (but also when not applying it!). The second explanatory case complied with both \textit{good practice} in the field of micro and nano manufacturing, and with the hypotheses of the TCL, showing how frequentist statistics can be useful in detecting and quantifying influence factors in the data set. The third case dealt successfully with a direct prove of a possible optimization of a typical surface characterization task. Additionally, drawbacks were also clearly indicated, mainly related 1) to the difficulty of complying with TCL hypotheses when measurements are long lasting, 2) to all the necessary cares needed for a correct implementation, 3) and to the possible overestimation of the traceability uncertainty contributor. Such inconveniences are not always avoidable, above all in an industrial environment. Eventually, the averaging of micrographs' height values stated as starting hypothesis unfolded favorably, resolving a complex correlation among pixels, even if the identification of the systematic behavior tended to be partial. On the contrary, the correction of the systematics pixel-wise was recognized inadequate, requiring completely new approaches possibly to be sought in the boosted perspective of {\sl Artificial Intelligence}, and by virtue of {\sl High Performance Computing}.

Therefore, the main reason of dismissal of a statistical exploitation of the information content lies in the aforesaid disadvantages of formulating a stochastic model. Conversely, the deterministic metrological characteristics' framework offers ease of application but requires heavy interplay with the measurements contents, which in turn can expose the operator to inaccuracies, and to an increase of the systematic behavior.

A possibility is foreseen in the use of self-calibration techniques \cite{evans}, where rediscovered methods need to be integrated by new ones, which are robust to noise, correlation and imperfection of the material measures. Self-calibration techniques are especially needed for the inspection of the linearity and perpendicularity of the axes in the instruments' reference system. For instance, the $x$ and $y$ mapping deviations are determined measuring the location of the center of mass of micro features in cross-grating material measures. Hence, they are heavily affected by any imperfection of the structured artifacts used, and depending on the result of the micrographs acquisitions.

To conclude, a twofold change of perspective is needed in manufacturing metrology: To make easier the access to complex mathematics in the fashion of software tools for industrial users is a major objective. The free-form structuring of surfaces and volumes requires a real 3D metrology, thus the traditional measurand must give way to the concept of distributed measurand.

\vspace{6pt} 

\ack{The author would like to thank prof. Giulio Barbato and dr. Gianfranco Genta at \textsl{Politecnico di Torino} for fruitful discussions, and for sharing valuable advises concerning statistical analysis and uncertainty evaluation. Author's gratitude also goes to prof. Hans N. Hansen at \textsl{Danmarks Tekniske Universitet---DTU} for his continuous support, guidance and inspiration.}


\appendix
\section{\label{appA}Dimensional and topographic metrology of micro structured tool inserts---supplemental data}

\begin{table}[ht]
\caption{\label{tab:A:height}Height of the steps $H{\rm~/\mu m}$ in the indicated measurand areas. A double value in the same column stands for repeated measurement. The values in parenthesis were excluded by the Chauvenet's criterion.}
\footnotesize
\begin{indented}
\lineup
\item[]
\begin{tabular}{@{}llllllll}
\br
Area & FV$^a$ $5\times$ & FV $10\times$ & LSC$^b$ $10\times$ & LSC $20\times$ & LSC $50\times$ & CSI$^c$ $20\times$ & CI$^d$\\
\mr
$A11$  &         & $162.1$ & $162.1$ & $160.0$ & $159.5, 161.0$   & $158.9$ & $162.6$\\
$A13$  &         & $163.9$ & $165.7$ & $163.5$ & $161.1, 161.4$   & $161.1$ & $162.3$\\
$A15$  &         & $167.3$ & $169.0$ & $167.6$ & $(159.2), 165.2$ & $163.7$ & $164.2$\\
$A25$  & $96.2$  &         & $95.9$  & $96.7$  &                  & $91.6$  & $97.3$\\
$A26$  & $83.9$  &         & $83.7$  & $84.8$  &                  & $85.2$  & $84.8$\\
$A27$  & $177.8$ &         & $177.1$ & $176.0$ &                  & $177.2$ & $177.2$\\
$A28$  & $85.6$  &         & $86.3$  & $84.5$  &                  & $83.1$  & $85.2$\\
$A29$  & $163.9$ &         & $161.5$ & $162.6$ &                  & $158.1$ & $163.1$\\
$A210$ & $80.9$  &         & $81.9$  & $80.8$  &                  & $78.6$  & $80.2$\\
\br
\end{tabular}\\
$^a$Focus Variation microscope; 
$^b$Laser Scanning Confocal microscope; 
$^c$Coherent Scanning Interferometry microscope; 
$^d$Contact Instrument.
\end{indented}
\end{table}
\normalsize

\begin{table}[ht]
\caption{\label{tab:A:sq}Root-mean-square values $Sq{\rm~/\mu m}$ of the indicated measurand areas. A double value in the same column stands for repeated measurements. The values in parenthesis were excluded by the Chauvenet's criterion.}
\footnotesize
\lineup
\begin{tabular}{@{}llllllllll}
\br
Area & FV$^a$ $5\times$ & FV $50\times$ & FV $100\times$ & LSC$^b$ $10\times$ & LSC $20\times$ & LSC $50\times$ & CSI$^c$ $20\times$ & CSI $50\times$ & CI$^d$\\
\mr
$A12$ &        & $\01.06$   & $\00.81$   &        & $1.47$ & $1.06, 0.89$ &        & $\00.87$   & $1.10$\\
$A14$ &        & $\00.80$   & $\00.76$   &        & $1.52$ & $0.80, 0.87$ &        & $\00.77$   & $0.98$\\
$A21$ & $5.25$ & $(3.56)$ & $(3.71)$ & $5.97$ & $5.51$ & $4.98$       & $5.23$ & $(6.52)$ & $5.19$\\
$A22$ & $1.91$ & $\00.44$   & $\00.36$   &        & $0.91$ & $0.77$       & $0.88$ & $\00.54$   & $0.78$\\
$A23$ & $1.69$ & $\00.44$   & $\00.41$   &        & $0.80$ & $0.53$       & $0.72$ & $\00.52$   & $0.49$\\
$A24$ & $1.58$ & $\00.40$   & $\00.24$   &        & $0.78$ & $0.45$       & $0.34$ & $\00.47$   & $0.44$\\
\br
\end{tabular}\\
$^a$Focus Variation microscope; 
$^b$Laser Scanning Confocal microscope; 
$^c$Coherent Scanning Interferometry microscope; 
$^d$Contact Instrument.
\end{table}
\normalsize

\begin{table}[ht]
\caption{\label{tab:A:devH}Deviations $devH{\rm~/\mu m}$ representing the values in table~\ref{tab:A:height} normalized (subtracted) to their respective areal averages, after applying the Chauvenet's criterion (excluded values in parenthesis).}
\footnotesize
\begin{indented}
\lineup
\item[]
\begin{tabular}{@{}lllllll}
\br
Area & FV$^a$ $5\times$ & FV $10\times$ & LSC$^b$ $10\times$ & LSC $20\times$ & LSC $50\times$ & CSI$^c$ $20\times$\\
\mr
$A11$  &          & $1.47$ & $1.53$   & $\-0.57$ & $\-1.08, 0.36$        & $\m\-1.69$\\
$A13$  &          & $1.13$ & $2.88$   & $0.68$   & $\-1.65,\m\m\-1.33$   & $\m\-1.71$\\
$A15$  &          & $1.98$ & $3.71$   & $2.22$   & $\-(6.15),\0\0\-0.10$ & $\m\-1.64$\\
$A25$  & $1.12$   &        & $0.78$   & $1.62$   &                       & $\m\-3.52$\\
$A26$  & $\-0.48$ &        & $\-0.68$ & $0.36$   &                       & $\m0.79$\\
$A27$  & $0.76$   &        & $0.13$   & $\-1.04$ &                       & $\m0.16$\\
$A28$  & $0.74$   &        & $1.41$   & $\-0.36$ &                       & $\m\-1.78$\\
$A29$  & $2.38$   &        & $0.002$  & $1.04$   &                       & $\m\-3.43$\\
$A210$ & $0.37$   &        & $1.31$   & $0.28$   &                       & $\m\-1.69$\\
\br
\end{tabular}\\
$^a$Focus Variation microscope; 
$^b$Laser Scanning Confocal microscope; 
$^c$Coherent Scanning Interferometry microscope; 
$^d$Contact Instrument.
\end{indented}
\end{table}
\normalsize

\begin{table}[ht]
\caption{\label{tab:A:devSq}Deviations $devSq{\rm~/\mu m}$ representing the values in table~\ref{tab:A:sq} normalized (subtracted) to their respective areal averages, after applying the Chauvenet's criterion (excluded values in parenthesis).}
\footnotesize
\lineup
\begin{tabular}{@{}lllllllll}
\br
Area & FV$^a$ $5\times$ & FV $50\times$ & FV $100\times$ & LSC$^b$ $10\times$ & LSC $20\times$ & LSC $50\times$ & CSI$^c$ $20\times$ & CSI $50\times$\\
\mr
$A12$ &        & $\00.04$     & $\0\-0.21$   &        & $0.44$ & $0.04,\0\0\-0.14$   &          & $\0\-0.16$\\
$A14$ &        & $\0\-0.12$   & $\0\-0.16$   &        & $0.60$ & $\-0.12,\0\0\-0.05$ &          & $\0\-0.15$\\
$A21$ & $0.16$ & $\-(1.54)$ & $\-(1.38)$ & $0.88$ & $0.42$ & $\-0.11$            & $0.14$   & $(1.43)$\\
$A22$ & $1.08$ & $\0\-0.39$   & $\0\-0.47$   &        & $0.08$ & $\-0.06$            & $0.05$   & $\0\-0.29$\\
$A23$ & $0.96$ & $\0\-0.29$   & $\0\-0.32$   &        & $0.07$ & $\-0.20$            & $\-0.01$ & $\0\-0.21$\\
$A24$ & $0.98$ & $\0\-0.21$   & $\0\-0.37$   &        & $0.17$ & $\-0.16$            & $\-0.27$ & $\0\-0.14$\\
\br
\end{tabular}\\
$^a$Focus Variation microscope; 
$^b$Laser Scanning Confocal microscope; 
$^c$Coherent Scanning Interferometry microscope; 
$^d$Contact Instrument.
\end{table}
\normalsize

\clearpage

\section{Correction of the systematic behavior vs time sequence---supplemental data}\label{appB}

\begin{table}[ht]
\caption{\label{tab:B:H}Step height measurements $H{\rm~/\mu m}$  in the measurand area \textsl{A13}.}
\footnotesize
\begin{indented}
\lineup
\item[]
\begin{tabular}{@{}lllll}
\br
$5\times$ & $10\times$ & $20\times$ & $50\times$ & $100\times$\\
Day 1     & Day 2      & Day 2      & Day 2      & Day 3\\
\mr
$163.7$&$163.4$&$162.1$&$162.3$&$162.3$\\
$163.6$&$163.4$&$162.8$&$162.5$&$161.8$\\
$163.5$&$163.2$&$162.2$&$162.2$&$161.8$\\
$164.1$&$163.4$&$162.3$&$162.6$&$161.3$\\
$163.9$&$163.8$&$162.6$&$162.9$&$163.3$\\
\br
\end{tabular}\\
\end{indented}
\end{table}
\normalsize

\begin{table}[ht]
\caption{\label{tab:B:Sq}Root-mean-square measurements $Sq{\rm~/\mu m}$  in the measurand area \textsl{A21}.}
\footnotesize
\begin{indented}
\lineup
\item[]
\begin{tabular}{@{}lllllll}
\br
$5\times$ & $10\times$ & $20\times$ & $20\times$ & $20\times$ & $50\times$ & $100\times$\\
Day 4     & Day 4      & Day 4      & Day 4      &   Day 5    & Day 6      & Day 6       \\
\mr
$5.66$&$5.22$&$5.05$&$5.10$&$5.08$&$5.07$&$5.56$\\
$5.46$&$5.06$&$5.22$&$5.07$&$5.06$&$5.06$&$5.59$\\
$5.52$&$5.13$&$5.08$&$5.08$&$5.09$&$5.08$&\\
$5.51$&$5.12$&$5.08$&$5.08$&$5.05$&      &\\
$5.46$&$5.07$&$5.15$&$5.06$&$5.05$&      &\\
\br
\end{tabular}\\
\end{indented}
\end{table}
\normalsize

\begin{table}[ht]
\caption{\label{tab:B:SqBand}Root-mean-square measurements $Sq{\rm~/\mu m}$  in the measurand area \textsl{A21}, re-sized at the same quantization level. The values in parentheses were excluded by the Chauvenet's criterion.}
\footnotesize
\begin{indented}
\lineup
\item[]
\begin{tabular}{@{}lllllll}
\br
$5\times$ & $10\times$ & $20\times$ & $20\times$ & $20\times$ & $50\times$ & $100\times$\\
Day 4     & Day 4      & Day 4      & Day 4      &   Day 5    & Day 6      & Day 6       \\
\mr
$5.27$&$\05.14$  &$5.17$&$5.30$&$5.23$&$5.40$&$(5.56)$\\
$5.31$&$\05.36$  &$5.14$&$5.21$&$5.24$&$5.39$&$(5.59)$\\
$5.40$&$\05.15$  &$5.18$&$5.21$&$5.26$&$5.39$&\\
$5.34$&$(6.61)$&$5.19$&$5.21$&$5.20$&      &\\
$5.28$&$(5.50)$&$5.29$&$5.24$&$5.22$&      &\\
\br
\end{tabular}\\
\end{indented}
\end{table}
\normalsize

\clearpage

\section{Correction of the systematic behavior in surface characterization---supplemental data}\label{appC}

\begin{table}[ht]
\caption{\label{tab:C:CI}CI measurements of the areal parameters $Sa$ and $Sq$ \cite{ISO25178-2}.}
\footnotesize
\begin{indented}
\lineup
\item[]
\begin{tabular}{@{}lllll}
\br
						     & $T1$ & $T2$  & $T3$  & $T4$\\
\mr
						     &$48.0$&$134.0$&$231.9$&$510.5$\\
						     &$47.5$&$131.6$&$232.4$&$510.3$\\
$Sa{\rm~/\mu m}$ &$47.6$&$133.3$&$232.6$&$510.5$\\
						     &$47.8$&$133.8$&$232.5$&$510.5$\\
\vspace{2mm}
						     &$47.4$&$133.3$&$232.9$&$510.3$\\
						     &$61.9$&$180.5$&$312.6$&$648.9$\\
						     &$61.2$&$177.1$&$313.2$&648.8$$\\
$Sq{\rm~/\mu m}$ &$61.2$&$179.5$&$313.5$&649.1$$\\
						     &$61.4$&$180.0$&$313.5$&$648.7$\\
\vspace{2mm}
						     &$60.9$&$179.3$&$313.9$&$648.9$\\
$U_{cal}{\rm~/\mu m}$&$10$	 &$10$	 &$10$	 &$24$\\
\br
\end{tabular}\\
\end{indented}
\end{table}
\normalsize

The following considerations where made with regards to the contact instrument (CI) measurements:
\begin{itemize}
\item Measurements of $Ra$ in the calibration certificate led to evaluate the capability of the reference stylus instrument to measure average height variations on a surface within a confidence interval (expanded uncertainty). For this reason, the uncertainty of CI related to $Ra$ measurements was consider compatible with measurements of $Sa$.
\item $Sa$ and $Sq$ are both amplitude parameters and strongly correlated (they measure the same quantity in different ways). The related uncertainties are normally in the same order of magnitude. Therefore, considering the same CI uncertainty for $Sq$ values, too, is an estimation of the uncertainty based on previous knowledge and experience \cite{GUM}.
\item The uncertainty contributor related to the calibration uncertainty is constant for all cases, related to both the correction of the systematic behavior and not. Hence, it is a common constant value and may affect the final evaluated uncertainty but not the investigation itself (correction of systematics). Nonetheless, a realistic final standard uncertainty for CI measurements was considered as $u_{CI}=$max$\left\{u_{cal},u_{repeatability}\right\}\cdot \sqrt{n_{input}}$, where $U_{cal}=k~u_{cal}$ ($k=2.2$ evaluated as inverse $t$-distribution on a confidence interval of 95~\% and degrees of freedom given by ${n_{input}-1}$), and, thus, spread on the averaged input multiplicity. In other words, being the CI repeatability small, it was

\begin{equation}\label{met_eq:uCI}
u_{CI}=u_{cal} \cdot \sqrt{n_{input}} 
\end{equation}

with $n_{input}=12$ repeated measurements stated in the calibration certificate.
\end{itemize}

\begin{table}[ht]
\caption{\label{tab:C:LSC}LSC$^a$ measurements of the areal parameter $Sq$. The values in parentheses were excluded by the Chauvenet's criterion.}
\footnotesize
\begin{indented}
\lineup
\item[]
\begin{tabular}{@{}llllll}
\br
&$(0,0)$&$(x,y)$&$(x,-y)$&$(-x,y)$&$(-x,-y)$\\
\mr
&$49.2$&$48.8$&$50.4$&$45.9$&$56.0$\\
&$48.9$&$49.5$&$49.9$&$45.0$&$46.3$\\
&$47.9$&$44.0$&$49.6$&$43.5$&$55.1$\\
&$49.2$&$48.5$&$49.9$&$41.1$&$54.2$\\
$T1$&$49.8$&$49.0$&$50.1$&$50.8$&$55.8$\\
&$49.8$&$49.4$&$44.5$&$44.3$&$56.3$\\
&$48.8$&$49.7$&$48.9$&$43.9$&$45.8$\\
&$49.2$&$44.0$&$42.8$&$39.4$&$53.2$\\
&$49.5$&$48.7$&$50.4$&$48.7$&$55.0$\\
\vspace{2mm}
&$48.5$&$43.1$&$43.6$&$42.9$&$55.2$\\
&$157.7$&$175.3$&$192.2$&$153.8$&$139.3$\\
&$157.2$&$176.2$&$195.1$&$152.0$&$138.4$\\
&$157.2$&$174.6$&$192.0$&$156.1$&$139.3$\\
&$159.1$&$175.6$&$191.6$&$153.6$&$135.9$\\
$T2$&$155.2$&$177.2$&$195.3$&$151.3$&$141.7$\\
&$166.5$&$179.4$&$193.8$&$158.7$&$142.5$\\
&$160.2$&$179.8$&$190.9$&$153.9$&$141.0$\\
&$163.1$&$174.7$&$195.7$&$152.5$&$142.3$\\
&$166.3$&$174.7$&$194.5$&$154.1$&$134.5$\\
\vspace{2mm}
&$157.9$&$174.6$&$193.1$&$153.0$&$141.0$\\
&$387.0$&$\0407.0$&$\0305.9$&$\0218.1$&$\0220.0$\\
&$392.9$&$\0404.0$&$(326.2)$&$\0223.7$&$\0225.0$\\
&$388.9$&$\0409.0$&$\0299.2$&$\0223.1$&$\0218.4$\\
&$387.1$&$\0410.8$&$\0305.3$&$(262.1)$&$\0224.0$\\
$T3$&$391.5$&$\0408.7$&$\0303.8$&$\0223.6$&$\0220.8$\\
&$385.9$&$\0409.1$&$\0308.6$&$\0221.1$&$\0221.2$\\
&$384.2$&$\0400.2$&$\0302.6$&$\0225.4$&$(258.2)$\\
&$385.3$&$\0411.1$&$\0308.9$&$\0221.8$&$\0218.9$\\
&$385.0$&$(343.4)$&$\0302.6$&$\0221.5$&$\0219.2$\\
\vspace{2mm}
&$388.3$&$\0402.8$&$\0318.9$&$\0228.0$&$\0225.9$\\
&$577.2$&$497.4$&$572.8$&$525.1$&$594.9$\\
&$574.0$&$494.2$&$570.3$&$522.8$&$597.6$\\
&$574.8$&$490.0$&$571.1$&$526.3$&$584.4$\\
&$579.0$&$498.4$&$573.3$&$524.4$&$593.9$\\
$T4$&$582.3$&$486.7$&$566.3$&$523.6$&$592.3$\\
&$577.4$&$484.0$&$568.4$&$523.0$&$595.3$\\
&$575.7$&$490.3$&$566.0$&$521.3$&$596.7$\\
&$573.2$&$483.8$&$573.6$&$528.0$&$604.0$\\
&$571.8$&$486.8$&$565.0$&$528.4$&$591.3$\\
&$576.2$&$499.7$&$567.9$&$524.6$&$592.0$\\
\br
\end{tabular}\\
$^a$Laser Scanning Confocal microscope
\end{indented}
\end{table}
\normalsize

\begin{table}[ht]
\caption{\label{tab:regression4}Results of the analysis of all the average values in the surfaces' sampling areas. Data corrected for the systematic behavior. $q$ is the slope of the model equation. $s_{regr,fit}$ is the standard deviation of the regression. $Sq_{corr}$ is the average of the root mean square values after correcting for the systematic behavior. $U$ is the expanded uncertainty (see \tref{tab:C:ucorr} below).}
\footnotesize
\begin{indented}
\lineup
\item[]
\begin{tabular}{@{}lllll}
\br
&$q$ /1&$s_{regr,fit}$ /nm&$Sq_{corr}$ /nm&$U$ /nm\\
\mr
\vspace{2mm}
$T1$&$0.790\pm0.009$&$1.8$&$48$&$28$\\
\vspace{2mm}
$T2$&$0.922\pm0.003$&$1.5$&$165$&$32$\\
\vspace{2mm}
$T3$&$0.972\pm0.002$&$1.8$&$309$&$35$\\
$T4$&$0.849\pm0.001$&$1.8$&$551$&$71$\\
\br
\end{tabular}\\
\end{indented}
\end{table}
\normalsize

\begin{table}[ht]
\caption{\label{tab:C:ucorr}Uncertainty budget of the measurements corrected for the systematic behavior, averaging in all areas. $|c_j|$ is the absolute value of the sensitivity coefficient of the generic uncertainty contributor $u(x_j)$ ($u_j(y)$ propagated contributors).}
\footnotesize
\begin{indented}
\lineup
\item[]
\begin{tabular}{@{}lllll}
\br
					 & 							 & $|c_j|$ & $u(x_j)$~/nm & $u_j(y)$~/nm\\
\mr
					 &$u_{CI}$			 &$0.8$ &$15.7$	&$12.4$\\
					 &$u_{repea,opt}$&$1.0$ &$1.7$	&$1.7$\\
$T1$			 &$u_{slope}$		 &$60.9$&$0.009$&$0.6$\\
\vspace{1.5mm}
					 &$u_{repr,fit}$ &$1.0$ &$1.8$	&$1.8$\\
					 &$k(95~\%)$     &			&				&$2.2$\\
\vspace{3mm}
					 &$U$						 &			&				&$28$\\
					 &$u_{CI}$			 &$0.9$ &$15.7$	&$14.5$\\
					 &$u_{repea,opt}$&$1.0$ &$1.5$	&1.5$$\\
$T2$			 &$u_{slope}$		&$177.0$&$0.003$&$0.5$\\
\vspace{1.5mm}
					 &$u_{repr,fit}$ &$1.0$ &$1.5$	&$1.5$\\
					 &$k(95~\%)$     &			&				&$2.2$\\
\vspace{3mm}
					 &$U$						 &			&				&$32$\\
					 &$u_{CI}$			 &$1.0$ &$15.7$	&$15.5$\\
					 &$u_{repea,opt}$&$1.0$ &$1.8$	&$1.8$\\
$T3$			 &$u_{slope}$		&$312.6$&$0.002$&$0.6$\\
\vspace{1.5mm}
					 &$u_{repr,fit}$ &$1.0$ &$1.8$	&$1.8$\\
					 &$k(95~\%)$     &			&				&$2.2$\\
\vspace{3mm}
					 &$U$						 &			&				&$35$\\
					 &$u_{CI}$			 &$0.8$ &$37.8$	&$32.1$\\
					 &$u_{repea,opt}$&$1.0$ &$1.8$	&1.8$$\\
$T4$			 &$u_{slope}$		&$648.7$&$0.001$&$0.6$\\
\vspace{1.5mm}
					 &$u_{repr,fit}$ &$1.0$ &$1.8$	&$1.8$\\
					 &$k(95~\%)$     &			&				&$2.2$\\
					 &$U$						 &			&				&$71$\\
\br
\end{tabular}\\
\end{indented}
\end{table}
\normalsize

\begin{table}[ht]
\caption{\label{tab:C:uncorr}Uncertainty budget of the non-corrected measurements, averaging in all areas ($|c_j|=1$, $u_j(y)$ propagated contributors).}
\footnotesize
\begin{indented}
\lineup
\item[]
\begin{tabular}{@{}lllll}
\br
						     & $T1$ & $T2$  & $T3$  & $T4$\\
\mr
$u_{CI}{\rm~/nm}$			 	&$15.7$ &$15.7$ &$15.7$ &$37.8$\\
$u_{repea,opt}{\rm~/nm}$&$1.7$&$1.5$&$1.8$&$1.8$\\
$u_{repea,CI}{\rm~/nm}$ &$0.4$&$1.2$&$0.5$&$0.1$\\
\vspace{1.5mm}
$u_{repr}{\rm~/nm}$		 	&$1.6$&$1.4$&$1.7$&$1.5$\\
$k(95~\%)$     					&$2.2$&$2.2$&$2.2$&$2.2$\\
$U{\rm~/nm}$						&$35$ &$35$ &$35$ &$83$\\
\br
\end{tabular}\\
\end{indented}
\end{table}

\end{document}